\documentclass[11pt]{article}
\newcommand{\blind}{1}

\usepackage[margin=1in]{geometry}

\usepackage{amsmath, amsthm, amssymb, bbm, bm} 

\usepackage[ruled]{algorithm2e}

\usepackage{comment}

\newtheorem{theorem}{Theorem}
\newtheorem{corollary}{Corollary}
\newtheorem{proposition}{Proposition}

\newtheorem{assumption}{Assumption}

\theoremstyle{definition}

\usepackage{graphicx} 
\usepackage{subcaption}
\graphicspath{{../../figures/}}

\usepackage[]{natbib}
\usepackage{setspace} 
\usepackage[utf8]{inputenc} 
\usepackage[T1]{fontenc}    
\usepackage{hyperref}  
\hypersetup{colorlinks,citecolor=blue,urlcolor=blue,linkcolor=blue}
\usepackage{url}            
\usepackage{booktabs}       
\usepackage{amsfonts}       
\usepackage{nicefrac}       
\usepackage{microtype}      

\begin{document}

\def\spacingset#1{\renewcommand{\baselinestretch}%
{#1}\small\normalsize} \spacingset{1}



\if1\blind
{
  \title{\bf Treatment Allocation with Strategic Agents}
  \author{Evan Munro
    \thanks{
I thank Mohammad Akbarpour, Susan Athey, Anirudha Balasubramanian, Martino Banchio, Alex Frankel, Guido Imbens, Ramesh Johari, Brad Ross, Stefan Wager, Gabriel Weintraub, Bob Wilson, and Kuang Xu for helpful comments and discussions.}\hspace{.2cm}\\
    Graduate School of Business, Stanford University\\
   }
  \maketitle
} \fi

\if0\blind
{
  \bigskip
  \bigskip
  \bigskip
  \begin{center}
    {\LARGE\bf  Treatment Allocation with Strategic Agents}
\end{center}
  \medskip
} \fi

\bigskip
\begin{abstract}
There is increasing interest in allocating treatments based on observed individual characteristics: examples include targeted marketing, individualized credit offers, and heterogeneous pricing. Treatment personalization introduces incentives for individuals to modify their behavior to obtain a better treatment.  Strategic behavior shifts the joint distribution of covariates and potential outcomes. The optimal rule without strategic behavior allocates treatments only to those with a positive Conditional Average Treatment Effect. With strategic behavior, we show that the optimal rule can involve randomization, allocating treatments with less than 100\% probability even to those who respond positively on average to the treatment. We propose a sequential experiment based on Bayesian Optimization that converges to the optimal treatment rule without parametric assumptions on individual strategic behavior. 
\end{abstract}

\noindent%
\vfill

\newpage
\spacingset{1.213}


\section{Introduction}
\label{sec:intro} 

The growing collection of individual-level data has increased the feasibility of personalizing treatments in a wide variety of settings. Treating individuals heterogeneously can improve outcomes compared to a uniform policy. \cite{rossi1996value} estimate a demand model to show that targeting consumers with different coupons depending on their purchase history can improve revenue compared to allocating the same coupon to everyone. Personalized medicine makes medical decisions based on individual characteristics, recognizing that treatments that fail for the average patient may be beneficial for certain subgroups of patients \citep{hamburg2010path}. Online lenders allocate credit based on conventional and unconventional individual characteristics like phone usage \citep{bjorkegren2019behavior}. 

When treatments have value to individuals, personalizing treatments introduces incentives for individuals to change their behavior used for targeting and receive a better treatment. In the coupon example, a budget-unconstrained and profit-maximizing seller would like to allocate a coupon to ``reluctant-buyers'', who will buy the product only if they receive a coupon. They would like to avoid giving the coupon to customers who would buy the product even without a coupon (``always-buyers''). The seller, however, cannot observe the buyer's type directly, so instead relies on proxies like online customer behavior. \citet{zhang2018price} examine the effects of a price promotion on Alibaba that targeted a subset of customers who left a product in their cart for 24 hours but had not checked out. They found that targeting these customers had an important unintended consequence which affected the revenue impact of the promotion: customers increasingly left items in their cart that they might otherwise have purchased immediately in an attempt to receive better prices. We will show that this type of strategic behavior affects the structure of the revenue-optimal policy that allocates coupons based on add-to-cart behavior. 

In this paper, we study the design of the optimal treatment rule that maximizes expected outcomes when agents are strategic.  Our model is a non-parametric Stackelberg game, where the planner first announces a treatment rule, which is a mapping from covariates to allocation probabilities for a binary treatment. Next, $n$ individuals report covariates (signals) strategically by maximizing a utility function that is unknown to the planner. Conditional on these strategically reported signals, binary treatments are allocated, and the corresponding potential outcome is realized for each individual. 

Agents may have value or disutility for the treatment. For a utility-maximizing agent that can change their behavior, it may be optimal to report a signal that leads to treatment with higher or lower probability. We do not impose a parametric model for agent utility functions. Instead, we require that the joint distribution of the agent's unobserved individual treatment effect and observed covariates varies smoothly with changes in the treatment rule. We show that this general assumption is satisfied by a variety of forms of agent behavior, including rational behavior where agents trade off value for the treatment and costs for changing their signal, as well as behavior where agents observe the treatment rule with error, or are not strategic at all. 
 
 The class of treatment rules considered is  simple. The planner is not budget constrained, and the only way they can interact with the agents is by announcing a treatment rule, so outside messages or transfers can't be used to align incentives. These modeling choices make the results in the paper comparable to those on the literature on treatment rules without strategic behavior or budget constraints \citep{manski2004statistical}. We consider a limited form of dynamics in the paper, where the planner can change the treatment rule over time, and observe outcomes for a batch of $n$ agents at each time step. We assume a myopic form of rationality for the agents; they behave in a way that maximizes their utility in the current period, but do not consider the impact of their choices on future agents. 

There is a growing literature on estimating prediction rules when agents strategically report covariates, including \citet{hardt2016strategic, dong2018strategic, perdomo2020performative, bjorkegren2020manipulation, miller2021outside}.  Optimal prediction rules when agents are strategic have a distinct structure compared to prediction functions in non-strategic settings, see \citet{frankel2019improving} and \citet{ball2020scoring}. The insights gained in the prediction setting do not apply directly to the causal setting. As described in \citet{athey2017beyond}, the problems of prediction and causal inference, while closely related, are distinct, which prevents us constructing optimal decision rules from an optimal prediction rule in general \citep{ascarza2018retention, bertsimas2020predictive}. In the coupon example, predicting which individuals are most likely to buy a product in response to receiving a coupon is not sufficient to construct the optimal allocation rule, since it allocates coupons to many always-buyers. The optimal rule allocates coupons to reluctant-buyers. Identifying reluctant-buyers requires a framework for counterfactual reasoning, since we cannot observe an individual's behavior with and without the coupon simultaneously.  The literature on policy learning derives welfare-maximizing treatment rules when covariates are exogenous \citep{manski2004statistical, bhattacharya2012inferring, kitagawa2018should,  hirano2009asymptotics, kallus2021minimax, athey2020policy}. In the absence of strategic behavior, the Conditional Average Treatment Effect is defined as the expected difference between treated and control outcomes for individuals with certain observed covariates. The optimal rule takes the form of a cutoff rule that assigns treatment with probability one to individuals with a positive CATE \citep{manski2004statistical}. What is missing from the literature, and which this paper seeks to address, is an analysis of treatment rules when there is strategic behavior. 

The first result of the paper is to derive a necessary condition that the optimal treatment rule under strategic behavior must satisfy. This condition relies on the existence of a generalized derivative of the planner's objective with respect to a directional change in the treatment allocation rule. When agents report their covariates strategically, each treatment rule leads to potentially distinct joint distribution of unobserved individual treatment effects and observed covariates. As a result, the average treatment effect conditional on a certain covariate value (the CATE) depends on the treatment rule implemented.   

We use this necessary condition to analyze how the structure of the optimal rule with strategic agents is distinct from the classical setting without strategic behavior. We define a cutoff rule as one that allocates treatments only to those who have a positive CATE, where the CATE is a functional of the distribution induced by the cutoff rule. When a cutoff rule exists, it is not always the optimal allocation rule. Instead, there exist natural settings where the optimal rule allocates the treatment to those with a positive CATE with probability less than one. Those with negative CATEs can receive the treatment with probability greater than zero. \citet{zhang2018price} report strategic behavior induced by a targeted price promotion. Our results imply that in their setting, a promotion that allocated coupons with 80\% probability to customers who left items in their cart might have improved revenue compared to the policy in the paper, which targeted with 100\% probability. The reason for this is that a cutoff rule can lead to an equilibrium where those with a negative Individual Treatment Effect (ITE) are incentivized to change their behavior to receive the coupon with a higher probability. When the planner introduces some randomization, they reduce the incentives for strategic behavior. The randomization leads to a joint distribution of treatment effects and covariates that makes it easier to distinguish between those with a negative and those with a positive ITE, improving outcomes. We show that a key parameter that determines the local optimality of a cutoff rule depends on the treatment effects of certain strategic individuals. A planner can decide whether they need to expand their policy search beyond cutoff rules by understanding what kind of individuals are incentivized to change their behavior to receive treatment with a higher probability. If it is those who have a positive treatment effect, then the planner can limit attention to cutoff rules. If it is those with a negative treatment effect, then the planner needs to expand their search space to include rules with some randomization.

In a general model with many multi-valued covariates, strategic behavior can be very complex and the objective function may be non-concave. To provide some additional insight into what kinds of strategic behavior lead to optimal allocation rules that have a cutoff compared to a randomization structure, we examine more closely a simple setting where agents report a single binary covariate used for targeting. We show how the structure of the optimal rule varies from a cutoff rule to a randomization rule as the sign and magnitude of the treatment effects of strategic individuals in the model changes. We provide two examples to further illustrate this point. In the first example inspired by \citet{rossi1996value}, a cutoff rule is not optimal, since it induces too much strategic behavior from individuals who would buy the product without a coupon but change their behavior to receive a coupon. The second example is allocating a product upgrade offer based on a measure of customer's expertise. In this model, incentives between agents and the planner are aligned and a cutoff rule is optimal.

In the absence of strategic behavior, it is straightforward to estimate the optimal rule using an A/B test to estimate CATEs, and assigning individuals with a positive estimated CATE to treatment, as described in \citet{manski2004statistical}. When there is strategic behavior and incentives for agents and the planner are not aligned, the policy search space is wider than without strategic behavior. The question that must be answered to solve for the optimal allocation rule is no longer ``which groups should be treated?'', but is instead ``with what probability should each group be treated?''. The optimality conditions that we derive for the optimal rule depends on behavioral parameters, such as how the marginal distribution of covariates shifts with a small change in the treatment rule, which can't be estimated using a traditional randomized experiment. The other contribution of the paper is to design a sequential experiment that allows the planner to learn the optimal treatment rule over time without any parametric assumptions on agent strategic behavior. We show that the problem of estimating the treatment rule can be cast as a zero-th order stochastic optimization problem. Then, a variety of techniques from the stochastic optimization literature can be used to estimate the optimal rule. With strategic behavior, it is not straightforward to verify that the planner's objective is concave in the allocation rule. As a result, we rely on a global optimization procedure based on Bayesian Optimization to estimate the optimal treatment allocation rule using sequential noisy function evaluations, which relies only on sufficient smoothness on the objective function \citep{snoek2012practical, letham2019constrained}. 

In the final section of the paper, we present an MTurk experiment that demonstrates that targeting a valuable treatment based on an observed signal induces a shift in the distribution of the signal conditional on the respondent's type. We use the data from this experiment to run a semi-synthetic simulation that demonstrates our proposed estimation procedure has average regret that decreases rapidly after a low number of noisy evaluations of the objective. 

\paragraph{Related Work} 

 In a principal-agent model, the principal designs a strategy, such a contract offered to an agent, in order to maximize the principal's outcomes by counteracting some informational asymmetry \citep{ross1973economic, myerson1982optimal}. Our model is a principal-agent model of screening, where agents have hidden information  \citep{spence2002signaling, stiglitz2002information}.  The agents have an unobserved type, which is their individual treatment effect. The planner would like to allocate treatment to those with a positive individual treatment effect, but observe a manipulable signal rather than the type. This model of hidden information is in contrast to a principal-agent model of moral hazard (hidden action), where the planner would like to incentivize certain actions (e.g. effort exertion) that are not observed directly by the planner \citep{akerlof1970market, grossman1981implicit}. 
 
 A screening model is one where the principal chooses their strategy first, and agents choose their signals knowing the principal's strategy.  Existing results on optimal screening contracts are not immediately translatable to our setting, since the principal strategy space is restricted to allocating a binary treatment based on an observed signal. Monetary transfers and other more complex instruments are not available to incentivize agents, as is common in the contract theory literature. Related work from \citet{yang2021costly} studies screening where the principal allocates monetary transfers and a good conditional on a non-price signal. In settings without monetary transfers, \citet{banerjee1997theory} examined how red-tape in bureaucracy can be used as a screening instrument.

There is a large related literature on strategic classification, which can also be considered as a principal-agent problem where the planner's instrument is a classification rule that predicts some outcome, rather than a treatment rule that impacts outcomes. Although prediction can be considered as a type of causal intervention \citep{miller2020strategic}, it is not possible to describe the treatment allocation problem with strategic agents as a classification problem, so the results from the strategic classification literature do not extend directly the causal setting. In the classification literature, there are models of hidden action where the planner uses a scoring rule to incentivize certain actions that are beneficial, but unobserved, see \citet{kleinberg2020classifiers} and \citet{haghtalab2021maximizing}. Most of the literature, however, are models of hidden information, where observed strategic actions are used to predict a label for an individual. The Stackelberg model in the prediction setting was introduced by \citet{hardt2016strategic}.  One strand of this literature examines algorithms that converge to the optimal prediction rule. Under assumptions on agent utility functions that lead to a convex optimization problem, \citet{dong2018strategic}, \citet{miller2021outside} and \citet{izzo2021learn} use well-known derivative-free convex optimization algorithms to converge to the optimal prediction rule over time.  \citet{bjorkegren2020manipulation} use a randomized experiment that varies the coefficients of a prediction function to estimate a parametric model of manipulation. \citet{ahmadi2021strategic} and \citet{chen2020learning} allow for agent strategic behavior that is less regular, so require new algorithms to estimate prediction rules that are robust to strategizing. Another strand of the strategic classification literature studies the convergence and optimality of algorithms that approximate the optimal strategy, such as repeated risk minimization \citep{perdomo2020performative, brown2020performative}. Rather than focusing on estimation of the optimal classification rule, there is work that derives the structure of the optimal rule. \citet{frankel2019improving} and \citet{ball2020scoring}  show that the optimal linear scoring rule underweights manipulable characteristics under a specific form of strategic behavior, where agents have a linear-quadratic utility specification. Under a different utility specification for agents, \citet{ahmadi2021strategic} finds that the optimal linear classifier has a threshold that is shifted by strategic behavior. We derive complementary results for treatment allocation. A form of underweighting appears, where the optimal treatment targeting rule with strategic agents can allocate treatments to those who respond positively to the treatment with less than 100\% probability. In a strategic classification problem with a fairness objective that takes the cost of strategic behavior into account, \citet{braverman2020role} also find that randomization can be optimal.

The literature on policy learning, which assumes agents are non-strategic and the distribution of observed covariates are fixed, has examined optimal treatment allocation rules in a growing variety of settings. The literature takes into account data in randomized \citep{manski2004statistical} and observational  settings \citep{athey2020policy, kallus2021minimax}, and examines the optimal rule under both finite sample and asymptotic frameworks \citep{hirano2009asymptotics}, as well as under budget constraints \citep{bhattacharya2012inferring} and other constraints on the allocation rule \citep{kitagawa2018should}. The theory in our paper examines the structure of the unconstrained optimal rule when agents are strategic in a setting otherwise similar to the initial papers in this literature, where treatments are binary and the objective is maximizing expected outcomes. 

The literature on robust treatment rules derives policies that are robust to perturbations in either the marginal distribution of covariates, or the joint distribution of covariates and potential outcomes \citep{mo2021learning, kallus2022doubly, adjaho2022externally}. Strategic behavior also results in a shift to the joint distribution of covariates and potential outcomes. The Stackelberg optimal rule maximizes outcomes when each treatment rule is associated with a single joint distribution of covariates and potential outcomes. This is in contrast to the distributionally robust optimization literature where the optimal treatment rule maximizes the minimum performance over a set of joint distributions of covariates and potential outcomes. This connection is explored in more detail at the end of Section \ref{sec:model}. 

The literature on incentivizing exploration considers an incentive issue that arises in a related, but distinct policy optimization problem \citep{frazier2014incentivizing, chen2018incentivizing, immorlica2020incentivizing, mansour2022bayesian}.  In this literature, myopic agents choose an action, based on imperfect information about rewards accumulated by previous actions. Since the agents make decisions myopically, and the rewards distribution is initially unknown by both the planner and the agents, agents do not explore all arms sufficiently to discover the long-term optimal policy. To better align short and long-term incentives, the planner can use payments to agents, recommendation systems, and other types of selective information disclosure.  Our model addresses a different issue leading to a mechanism design problem in policy optimization, which is information asymmetry.  Agents are fully informed about their own utility function, and depending on their value for the treatment and the treatment allocation function, choose a signal to maximize this reward. The planner, however, has a distinct reward function that depends on the agent's type, which the planner does not know. The planner must learn how to use the manipulable signal to target treatments to agents with a certain hidden type. 

There is a large literature on experimental designs that go beyond a basic A/B test, see \citet{cornfield1978randomization} and \citet{athey2017econometrics}  for a discussion of stratified and cluster-randomized experiments in settings without interference. Recently, there is increasing interest in using new experimental designs to estimate causal quantities of interest under interference between treated and control units \citep{hudgens2008toward, savje2021average}. There is a large literature on designs when there is network or spatial structure in the data,  such as clustering or partial interference, an incomplete review includes  \citet{tchetgen2012causal, ugander2013graph, baird2018optimal, offer2021experimentation} and \citet{ vazquez2017identification}. In two-sided markets, the structure of the market leads to a variety of experimental designs based on cluster or individual-level randomization, see \citet{harshaw2021design, bajari2021multiple, munro2021treatment, liao2023statistical} and \citet{viviano2020experimental}. This paper shows that under strategic effects, without the presence of interference, new experiment designs are needed to learn how to target treatments effectively. 

\section{Models of Treatment Allocation}
\label{sec:model} 

\subsection{Treatment Allocation with Exogenous Covariates}
\label{sec:standard} 
We start by discussing the classical setting of \citet{manski2004statistical}. Each of $i \in 1, \ldots, n$ individuals has real-valued potential outcomes $\theta_i = \{ Y_i(1), Y_i(0) \}$ and covariates $X_i \in \mathcal X$. The covariates may be discrete or continuous-valued. The covariates and potential outcomes are random variables that are independent and identically distributed according to a joint distribution, $F_{x, \theta}$. 

The planner has control over the assignment of a binary treatment $W_i \in \{0, 1\}$. Let $\tau_{\theta}(\theta_i)= Y_i(1) - Y_i(0)$. With full information, an outcome-maximizing planner would assign treatment to anyone with $\tau_{\theta}(\theta_i) > 0$ and withhold treatment from those with $\tau_{\theta}(\theta_i) \leq 0$. However, only $X_i$ is observed before treatments are allocated, so the planner chooses an allocation rule that assigns treatment conditional on $X_i$, rather than $\theta_i$. The allocation proceeds as follows: 

\begin{enumerate} 
\item The planner specifies  $\pi: \mathcal X \rightarrow [0, 1]$, where $\pi(x) = Pr(W_i = 1 | X_i =x)$. 
\item  A binary treatment $W_i$ is sampled from $\mbox{Bernoulli}(\pi(X_i))$.  
\item  The observed outcome is $Y_i = Y_i(W_i)$. 
\end{enumerate} 

The planner chooses $\pi: \mathcal X \rightarrow [0, 1]$ to maximize the expected outcomes $\mathbb E[Y_i(W_i)]$. Let $\tau_x(x) = \mathbb E[Y_i(1) - Y_i(0) | X_i = x]$ be the CATE, the average treatment effect among individuals who have covariate value $x$. We can expand the objective to show that $\pi(x)$ enters the objective linearly, and this leads to a simple optimal treatment rule that takes a cutoff form.  Using Bayes' rule, 

\begin{equation} \label{eqn:exog} \mathbb E[Y_i(W_i)] = \int \left ( \pi(x) \mathbb E[Y_i(1) | X_i = x] + ( 1- \pi(x)) \mathbb E[Y_i(0) | X_i =x ]\right ) dF_x(x).  \end{equation}

\begin{proposition} 
\label{prop:cesgood}
Assume that potential outcomes are bounded. For any $x \in \mathcal X$ with positive support, the treatment allocation rule that maximizes (\ref{eqn:exog}) is defined by $ \pi^0(x) = \mathbbm{1} ( \tau_x(x) > 0)$. 
\end{proposition} 

The proof of this proposition is straightforward and appears in different forms in the literature; see for example Setting 3 of \citet{vanderweele2019selecting}. In order to estimate this rule based on a finite sample of data, we require only an estimate of $\tau_x(x)$, which can be constructed using data from a Bernoulli randomized experiment. For example, in a setting with discrete $X_i$, a consistent estimator is: 
\[ \hat \tau_x(x) = \frac{\sum \limits_{i=1}^n \mathbbm{1}(X_i = x, W_i = 1) Y_i }{ \sum \limits_{i=1}^n \mathbbm{1}(X_i = x, W_i = 1) } -  \frac{\sum \limits_{i=1}^n \mathbbm{1}(X_i = x, W_i = 0) Y_i }{ \sum \limits_{i=1}^n \mathbbm{1}(X_i = x, W_i = 0) }. \]

The choice $\hat { \pi}^0(x) = \mathbbm {1} ( \hat \tau_x(x) >0)$ for $x \in \mathcal {X}$ is the Conditional Empirical Success Rule of \citet{manski2004statistical}. In the next section, we will show how under strategic behavior then $\pi(x)$ no longer enters the planner's objective linearly, leading to a more complex structure for the optimal rule. 

\subsection{Treatment Allocation with Strategic Agents} 

We now introduce our main setting, where $X_i$ is chosen strategically by each individual in response to the treatment allocation rule, rather than drawn randomly.  Each of $i \in \{ 1, \ldots, n \}$ individuals have a multidimensional type $\theta_i  = \{Y_i(1), Y_i(0), S_i \}  \in \Theta$  that is independently drawn from distribution $F_{\theta}$.  $\{Y_i(1), Y_i(0) \} \in \mathcal Y \times \mathcal Y$ are bounded and real-valued potential outcomes. $\bm S_i \in \mathbb R^k$ governs how an agent responds strategically to a treatment allocation rule. The treatment allocation rule is a mapping $\pi: \mathcal X \rightarrow [0, 1]$ from the space of signals (or covariates) $\mathcal X$ to an individual's  treatment probability. The space of possible treatment allocation rules, $\Omega$, is a vector space, such as all continuous functions from $\mathcal X$ to $[0, 1]$. 

The individual treatment effect is defined as before as $\tau_{\theta}: \Theta \rightarrow \mathbb R$, where $\tau_i =  \tau_{\theta} (\theta_i) =  Y_i(1) - Y_i(0)$. Individual strategic behavior is describe by a choice function, which is a measurable function $x: \Theta \times \Omega \rightarrow \mathcal X$ that maps a given treatment allocation rule and the agent's type to a choice of signal, $X_i^{\pi}  = x(\pi, \theta_i)$. Fixing a given treatment allocation rule $\pi$, the mappings $\tau_{\theta}(\theta)$ and $x(\pi, \theta)$ induce a joint distribution of the random variables $X^{\pi}_i$ and $\tau_i$, which we denote $F^{\pi}_{x, \tau}$. We can provide a microeconomic foundation for the choice function as the result of maximization of a type-specific utility function $u: \Omega \times \mathcal X \times \Theta \rightarrow \mathbb R$: 
\[ x(\pi, \theta_i) = \arg \max \limits_x u(x, \pi, \theta_i) \]

Utility functions can vary for different individuals, since the strategic-type is individual-specific. Let $\pi_0$ denote the control treatment rule, where $\pi(x) = 0$ for every $x \in \mathcal X$. We can define the agents' default choice of signal as $X_i^0 = \arg \max \limits_x u(x, \pi_0, \theta_i)$. An individual's utility function determines how much they value a treatment, as well as how costly it is to them to change their default preferred behavior $X_i^0$. This determines how an individual with a certain type responds to a given treatment allocation rule.  If for some $\pi' \neq \pi_0$, $\arg \max_x u(x, \pi', \theta_i) \neq  X_i^0$ for some non-zero measure of $\theta_i$, then there is strategic behavior, and incentives introduced by the treatment rule will shift the joint distribution of $\tau_i$ and $X_i^{\pi}$. 

We do not impose a specific parametric form for the utility function. However, agent behavior cannot be completely arbitrary; Theorem \ref{thm:first} requires an aggregate smoothness condition on $F^{\pi}_{x, \tau}$. In Proposition \ref{prop:utility}, we introduce some examples of utility functions that meet this regularity condition and can capture a variety of forms of agent preferences in a setting with binary covariates. These examples include agents that have simple additive preferences for the treatment and certain signals, agents that have more complex non-linear preferences for the treatment and certain signals, as well as agents that observe $\pi$ with some error. For now, we simply assume that the choice function $x(\pi, \theta_i)$ exists and is measurable. 

We can now describe the treatment allocation  procedure as a Stackelberg game:  
\begin{enumerate} 
\item The planner specifies  $\pi: \mathcal X \rightarrow [0, 1]$, with $Pr(W_i = 1 | X_i = x) = \pi(x)$. 
\item For $ i \in [n]$, agent $i$ reports covariates $X_i^{\pi} = x(\pi, \theta_i)$. 
\item For $i \in [n]$, $W_i$ is sampled from $\pi(X_i^{\pi})$. 
\item The outcome $Y_i = Y_i(W_i)$ is observed\footnote{ In this section, where we characterize the structure of the optimal rule, we assume that potential outcomes do not depend on the treatment allocation rule. However, the estimation procedure given in Section \ref{sec:est} is robust to settings where potential outcomes also depend on $\pi$.}.
\end{enumerate}

The key difference in this model compared to the model in the previous section is the introduction of Step 2, where $X_i$ is chosen strategically based on the function $x(\pi, \theta_i)$. The joint distribution of the observed signal and the unobserved individual treatment effect$, F^{\pi}_{x, \tau}$ now depends on the treatment rule $\pi$. This makes the planner's objective function depend on $\pi$ in a non-linear way. 

The objective for the planner is to choose an allocation rule that maximize expected outcomes $\mathbb E[Y_i(W_i)]$. Maximizing $\mathbb E[Y_i(W_i)]$ with respect to the function $\pi$ is equivalent to maximizing $ V(\pi) = \mathbb E[\pi(X^{\pi}_i)( Y_i(1) - Y_i(0))]$.  We define the optimal Stackelberg rule as the treatment allocation rule that maximizes $V(\pi)$ : $\pi^* = \arg \max \limits_{\pi} V(\pi).$ Using Bayes' rule, we can write \[V(\pi) = \arg \max \limits_{\pi}  \int  \pi(x) \mathbb E[Y_i(1) - Y_i(0) | X^{\pi}_i = x]  dF^{\pi}_x(x).\]
Compared to Equation \ref{eqn:exog}, the marginal distribution of the signal and the distribution of potential outcomes conditional on the signal are no longer fixed, but instead depend on $\pi$. 


\subsection{Characterizing the Optimal Treatment Rule} 
\label{sec:theory} 
The next step is to show how this can affect the structure of the optimal treatment rule. In the standard setting without strategic behavior, learning the conditional average treatment effect $\tau_x(x)$ was sufficient for constructing the optimal rule. However, in the Stackelberg setting, the population average treatment effect conditional on a given signal value is no longer a fixed quantity. Since the joint distribution of $\tau_i$ and $X^{\pi}_i$ now depends on $\pi$, the CATE is now indexed by the treatment allocation rule: 
\begin{equation*} 
\begin{split}
 \tau_x(\pi, x) & = \int \tau dF^{\pi}_{x, \tau} ( \tau | x) \\ 
& = \mathbb E[Y_i(1) - Y_i(0)| X_i^{\pi} = x ]. 
\end{split} 
\end{equation*} 

Theorem \ref{thm:first} derives a necessary condition that an optimal rule must satisfy. This condition depends on this new version of the CATE, and an additional component $s(\pi, x)$ which captures how changes in $\pi(\cdot)$ affect the joint distribution of $\tau_i$ and $X_i^{\pi}$. 




 
\begin{theorem} \label{thm:first}
Assume that the following directional derivative exists,  for all directions $h: \mathcal X \rightarrow \mathbb R$: 
\begin{equation} \label{eqn:first} \lim \limits_{ \delta \rightarrow 0 } \int  \tau \cdot  \pi(x)  \frac{dF^{\pi + \delta h} _{x, \tau}(x, \tau)  - dF^{\pi}_{x, \tau} (x, \tau ) }{\delta} = \int h(x) s(\pi, x) d \mu (x), \end{equation}
where $\mu(x)$ is the counting measure for discrete $x$ and the Lebesgue measure for continuous $x$. 

Then, the Gateaux derivative of the objective $V(\pi)$ in the direction $h(x)$ is 
\[ \partial V(\pi ; h) = \int h(x) [f_x(\pi, x) \tau_x( \pi, x) + s(\pi, x) ]d \mu (x), \]

where $f_x(\pi^*, x)$ is the probability density for continuous $x$ and the probability mass function for discrete $x$. 

Furthermore, any optimal rule $\pi^* \in \arg \max V(\pi)$ must satisfy the following conditions almost surely: 
\begin{enumerate} 
\item $\pi^*(x) = 0$ and $f_x(\pi^*, x) \tau_x( \pi^*, x) + s(\pi^*, x) \leq 0$, or
\item $\pi^*(x) = 1$ and $f_x(\pi^*, x) \tau_x( \pi^*, x) + s(\pi^*, x) \geq 0$, or 
\item $0 \leq \pi^*(x) \leq 1$ and $f_x(\pi^*, x) \tau_x( \pi^*, x) + s(\pi^*, x) = 0. $ 
\end{enumerate}

\end{theorem}

The proof of this theorem, in Appendix \ref{pf:first}, relies first on deriving the directional derivative of $V(\pi)$ in $\pi$. Then, the result  follows from a necessary condition for a local maximum of $V(\pi)$ from Theorem 2 of Chapter 7 of \citet{luenberger1969optimization}. If there is no strategic behavior and the joint distribution of $X_i$ and $\tau_i$ does not shift with changes to $\pi$, then $s(\pi, x) = 0$ and we are back in the setting of Proposition \ref{prop:cesgood}. If the CATE is positive, then the optimal rule always has $\pi^*(x)=1$.  But with strategic behavior, then $f_x(\pi, x) \tau_x( \pi, x) + s(\pi, x) $ depends also on the sign and magnitude of $s(\pi, x)$. The sign and magnitude of $s(\pi, x)$ depends on how the joint distribution of the signal $X_i$ and the unobserved treatment effect $\tau_i$ shifts with directional changes to $\pi$. If $s(\pi, x)$ is negative when $\tau_x(\pi^*, x)$ is positive, then it is possible that $0 < \pi^*(x) < 1$ and the optimal rule involves randomization of the treatment. 

Before analyzing this implication further, we first shed some light on what kind of agent behavior meets Condition \ref{eqn:first}  in the Theorem, and derive an explicit characterization of $s(\pi, x)$. We introduced the concept of an agent-specific utility function to rationalize agents behaving strategically in response to a given treatment allocation rule. In previous work on strategic classification, it is common to choose a single parametric form for agent utility, and prove results that hold for this utility function, see \citet{dong2018strategic} and \citet{frankel2019improving}. Recent work by \citet{chen2020learning} and \citet{ahmadi2021strategic} make less restrictive assumptions on agent behavior. The characterization of the optimal rule in Theorem \ref{thm:first} hold under a variety of parametric assumptions on agent behavior, as long as Condition \ref{eqn:first} holds. Proposition \ref{prop:utility} provides some examples of utility functions for which Condition \ref{eqn:first} holds. We focus on the binary setting for this proposition, which leads to simple utility and choice functions. 

\begin{proposition} \label{prop:utility} 
Let $X_i \in \{A, B\}$ be a binary covariate. Four different utility functions are defined below: 
\begin{enumerate} 
\item Non-Strategic Agents: $u_0(x, \pi, \theta_i) = - C_{ix}$ 
\item Boundary Crossing Model: $u_1(x,\pi, \theta_i)  = V_i \pi(x) - C_{ix}$ 
\item Imperfectly Informed Agents: $u_2(x, \pi, \theta_i)  = V_i (\pi(x) - \mathbbm{1}(x=A)( 1- \pi(x)) - C_{ix}$ 
\item Risk Averse Agents: $u_3(x, \pi, \theta_i) = \pi(x) V_i - r\cdot \pi(x) (1 - \pi(x))  - C_{ix}$
\end{enumerate} 

  Assume that each element of $\theta_i$ is bounded. Let $Z_i = 1$ if there exists $\pi, \pi' \in \Omega$ such that $\arg \max_{x} u(x, \pi, \theta_i) \neq \arg \max_{x} u(x, \pi', \theta_i)$. This is an indicator if an agent is strategic. For each $\tau \in \mathcal T$ with $Pr(Z_i = 1 | \tau_i = \tau) > 0$, assume  the distribution of the random variable $(C_{iA} - C_{iB}) | \tau, Z_i =1 $ is absolutely continuous. Then, the joint distribution of $X_i^{\pi}$ and $\tau_i$ induced by maximization of $u_k$ meets Condition (\ref{eqn:first}) for $k \in \{0, 1, 2, 3\}$. 

The following utility function, however, leads to agent behavior that does not meet the conditions of Theorem \ref{thm:first}. 

\begin{enumerate} 
\item $ u(x, \pi, \theta_i) = \mathbbm{1}(\pi(x) > 0.5) - C_{ix}$
\end{enumerate} 

\end{proposition} 

The proof of this Proposition is in Appendix \ref{pf:utility}. In the first example, the agent does not value the treatment, so the treatment allocation rule doesn't enter into the agent utility function. The agent always chooses the covariates with the smallest $C_{ix}$, and we are back in the non-strategic setting. In the next example, agents have value $V_i$ for the treatment, and a cost of choosing each signal $C_{ix}$. This boundary-crossing model leads to a simple form for the choice function for those with $V_i >0$: 
 \[ \mathbbm{1}(X_i = A) = \mathbbm{1}\left ( \pi(A) - \pi(B) \geq \frac{C_{iA} - C_{iB}}{V_i} \right).\]

In the third example, agents are imperfectly informed about the treatment rule. When choosing a signal, the agents always respond as if $\pi(A) =1$, even if it is less than 1, but observe $\pi(B)$ correctly. This will lead to a different optimal treatment rule than the setting where agents respond to $\pi(A)$ and $\pi(B)$ correctly. In the fourth example, agent utility depends on the treatment allocation rule in a non-linear way. They value receipt of the treatment, but penalize signals that receive treatment non-deterministically. Settings where agents have behavior described by any of these utility functions (or a mixture of them) lead to a joint distribution of $X_i^{\pi}$ and $\tau_i$ that varies smoothly with $\pi$. This list is by no means comprehensive, and it is possible to derive a variety of other utility functions that also meet this condition. 

In contrast, the last example given in the Proposition leads to a joint distribution of $X_i^{\pi}$ and $\tau_i$ that is discontinuous in $\pi$. Agents only recognize the value of the treatment when $\pi(x) > 0.5$. This means that the marginal distribution of $X_i^{\pi}$ has a discontinuous jump at $\pi(x) =0.5$.

In the first four examples of Proposition \ref{prop:utility} with binary covariates, agents have reporting behavior which is discontinuous, but the aggregate distribution of $X_i^{\pi}$ is still smooth in $\pi$, and agent behavior meets Equation \ref{eqn:first} of Theorem \ref{thm:first}. For settings with general discrete covariates, Proposition \ref{prop:suf} introduces a simple condition on the joint distribution of $\tau_i$ and $X_i^{\pi}$ that is sufficient for Equation \ref{eqn:first} of Theorem \ref{thm:first} to hold. Proposition \ref{prop:suf}  requires that the conditional probability mass function of $X_i^{\pi}$ is differentiable in $\pi(x)$. This restriction also leads to an analytical form for $s(\pi, x)$, which is helpful for understanding in what kind of scenarios that $0 < \pi^*(x) < 1$. 

\begin{proposition} \label{prop:suf} 
Let $X^{\pi}_i \in \mathcal X$ be discrete, with $| \mathcal X| = d$. Let  $\tau_i\in \mathcal T$. Assume that for all $\pi \in \Omega$, $t \in \mathcal T$,  and each $x', x \in \mathcal X \times \mathcal X$, $Pr(X_i^{\pi} = x | \tau_i = t )$ is differentiable in $\pi(x')$. Then,

\begin{equation} \label{eqn:sx} s(\pi, x) = \sum_{x' \neq x } [\pi(x') - \pi(x) ]\mathbb E \left [ \tau_i \cdot \frac{\partial Pr(X_i^{\pi} = x' | \tau_i)}{\partial \pi(x) } \right]   \end{equation} 
\end{proposition} 

The proof of this result is in Appendix \ref{pf:suf}. With discrete covariates, then the treatment allocation function can be represented by a vector in $[0, 1]^d$. The directional derivative of the objective can be written as $\partial V(\pi; h) = \sum \limits_{x \in \mathcal X} h(x) \frac{\partial V(\pi)}{\partial \pi (x)} $, where \[\frac{\partial V(\pi)}{\partial \pi (x)} =   [f_x(\pi, x) \tau_x( \pi, x) + s(\pi, x)].\] 

In Appendix \ref{pf:ext}, we provide a result that extends this Proposition also to continuous $x$, relying on the existence of a directional derivative of the conditional probability density function rather than a standard derivative of the conditional probability mass function. We also note that the differentiability condition from Proposition \ref{prop:suf} is sufficient, but not necessary for the existence of $s(\pi, x)$ defined in Theorem \ref{thm:first}. In Appendix \ref{pf:comment} we include an example where the probability mass function of $X_i$ conditional on $\tau_i$ is not differentiable in $\pi(x)$, but the function $s(\pi, x)$ still exists.

Proposition \ref{prop:suf} indicates that the sign and magnitude of $s(\pi, x)$ depends what type of agent is incentivized to change their signal in response to a marginal change in $\pi(x)$. Let's imagine that for a given treatment rule with $\pi(x) <1$,  individuals who report $X_i = x$ have a positive treatment effect on average ($\mathbb E[Y_i(1) - Y_i(0) | X_i^{\pi} = x] > 0$). The planner can directly increase $V(\pi)$ by increasing $\pi(x)$, which raises the probability that $Y_i(1)$ is observed rather than $Y_i(0)$ for agents reporting $X_i^{\pi}=x$. But, raising $\pi(x)$ also has an indirect impact in that it affects which agents report $X_i^{\pi} =x$, since when agents are strategic, we must also consider their incentives to report different signals. This indirect impact is captured by $s(\pi, x)$. An agent can have a positive utility for the treatment, and may respond to an increase in $\pi(x)$ by reporting $x$ instead of $x'$. Since agent utilities are in general distinct from agent outcomes defined by the planner, agents who value the treatment most do not always have a positive treatment effect. 

When it is  agents with negative treatment effects that  are incentivized to switch from reporting $x'$ to $x$ when $\pi(x') < \pi(x)$, then for those agents, $Y_i(1)$ is observed with higher probability instead of $Y_i(0)$. This has a negative impact on $V(\pi)$, which is captured by $s(\pi, x)$ being negative.  This means that even for a signal where the conditional average treatment effect $\tau_x(\pi, x)$ is positive, the derivative of $V(\pi)$ with respect to $\pi(x)$ can be negative if $s(\pi, x)$ is negative and large. This will impact the structure of the optimal rule. 

Now that we have some intuition on what determines $s(\pi, x)$, we return to the setting where $X_i \in \mathcal X$ can be either discrete or continuous. A natural conjecture for a targeting rule that might perform well is an extension of the cutoff rule from Proposition \ref{prop:cesgood}. The cutoff rule with strategic agents is one that meets the condition  
\begin{equation}
\label{eq:cutoff}
\pi^c(x) = \mathbbm{1} ( \tau_x(\pi^c, x ) >0 ). 
 \end{equation} 
This rule allocates treatments only to individual who have a positive CATE, where the CATE is defined using the distribution $F^{\pi^c}_{\tau, x}$ . A cutoff rule meeting this fixed point condition is not guaranteed to exist. In settings where it does exist, we show in Corollary \ref{cor:tilde} that the optimal allocation rule sometimes has this form, but in other cases does not. 

\begin{corollary}
\label{cor:tilde}
Assume that a cutoff rule of the form \[ \pi^c(x) = \mathbbm{1} (\tau_x(\pi^c, x) >0) \]  exists. If there is any $\tilde x \in \mathcal X$ such that  $\mbox{sgn}( s(\pi^c, \tilde x)) \neq \mbox{sgn} ( \tau_x(\pi^c, \tilde x))$ and $|s (\pi^c, \tilde x)| > | f(\pi^c, x) \tau_x(\pi^c, \tilde x) |$, then $\pi^* \neq \pi^c$ and the optimal rule does not have a cutoff form. If no such $\tilde x$ exists, and $V(\pi)$ is concave in $\pi$, then the cutoff rule is optimal even in the presence of strategic behavior. 
\end{corollary}

We can no longer guarantee that a cutoff rule is optimal if the strategic effect $s(\pi^c, x)$ is large enough and of opposite sign to the CATE, $\tau_x(\pi^c, x)$. Under the conditions described in Corollary \ref{cor:tilde}, the optimal rule can be an interior solution, where for certain values of $x \in \mathcal X$, we induce some randomization, so that $0 < \pi^*(x) < 1$.

We end this section with a simple example with binary covariates that illustrates Corollary \ref{cor:tilde}. In this example, described in Assumption \ref{ass:strat}, there are three types of agents, who each have a utility function following $u_1(x, \pi, \theta_i)$ from Proposition \ref{prop:utility}. We restrict strategic behavior to take the form of selection from $B$ into $A$ by a some proportion of agents \citep{heckman2005structural}.

\begin{assumption} \label{ass:strat} $\theta_i = \{ V_i, \tau_i, C_{iA}, C_{iB} \}$ and the utility function is $u(x, \pi, \theta_i) =  V_i \pi(x) - C_{ix}$, with $V_i \geq 0$. 
\begin{itemize} 
\item With probability $\rho_H$, agents have $\tau_i = H >0$, $C_{iA} < C_{iB}$, and $V_i < C_{iB} - C_{iA}$. They always report $X_i^{\pi} = A$.
\item With probability $\rho_M$, agents have $\tau_i = M$, $V_M >0$, $C_{iA} > 0 = C_{iB}$ and $\{C_{iA} \}$ has an absolutely continuous distribution. They are strategic and $X_i^{\pi} = A$ when $\pi(A) - \pi(B) > \frac{C_{iA}}{V_M}$. 
\item With probability $1- \rho_H - \rho_M$, agents have $\tau_i = L < 0$, $C_{iA} > C_{iB}$, and $V_i < C_{iA} - C_{iB}$. They are not strategic and always report $X_i^{\pi} = B$.
\end{itemize} 
\end{assumption}

Agents with $\tau_i = H$ have a positive individual treatment effect, and have $C_{iA} < C_{iB}$ such that they always report $X_i = A$. Agents with $\tau_i = L$ have a negative individual treatment effect and have $C_{iB} < C_{iA}$ so that they always report $X_{i} = B$, no matter the treatment rule. There are a proportion $\rho_M$ of agents who have $\tau_i = M$. In the absence of incentives otherwise, they will report $B$. However, if $\pi(A) > \pi(B)$, some of these agents may report $A$ instead. When $\rho_M = 0$, we are back in the non-strategic setting, and the cutoff rule $\pi(A) = 1$ and $\pi(B) = 0$ is optimal, since $\tau_x(A) >0$ and $\tau_x(B) <0$. When $\rho_M >0$, then the strategic behavior of those with $\tau_i =M$ can affect the structure of the optimal rule. Under this model, Condition \ref{eqn:first} of Theorem \ref{eqn:first} is met, with 
\begin{equation}\label{eqn:binary}  s(\pi, A) = [\pi(A) - \pi(B) ]\cdot  \rho_M \cdot  M \cdot V_M \cdot f^M_{cA}\Big(V_M[\pi(A) - \pi(B)] 
\Big)\end{equation} 
where $f^M_{cA}(c)$ is the density of $C_{iA} | \tau_i = M$, see Appendix \ref{pf:binary} for a derivation.

\begin{figure}
\centering
\includegraphics[width = 0.7\textwidth]{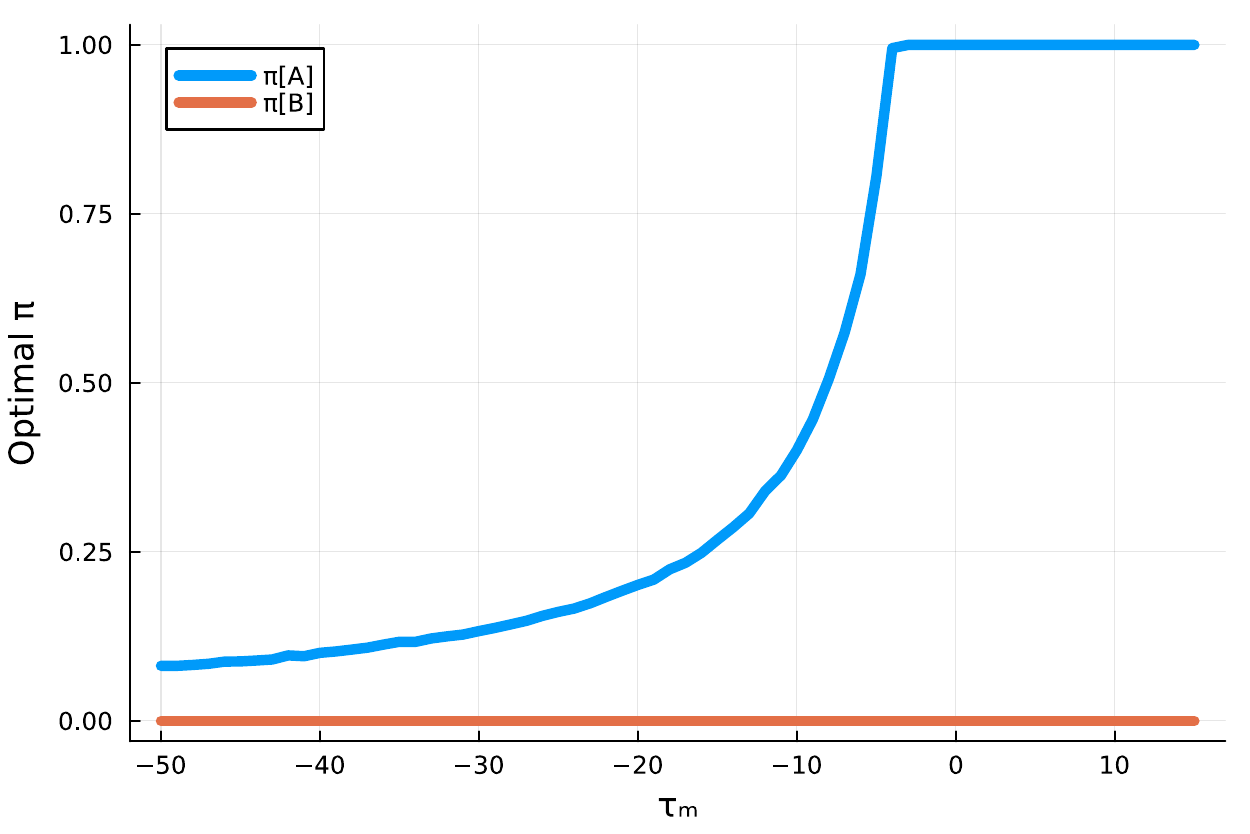} 
\caption{The optimal treatment allocation rule following Assumption \ref{ass:strat} as the value of $\tau_M = M$ varies. $H = 4.0$ and $L=-5.0$. $\rho_M = \rho_H = \frac{1}{3}$, $C_{iA} | (\tau_i =M )\sim \mbox{Uniform}(0, 2)$, and $C_{iB} | ( \tau_i = M )= 0$. \label{fig:cutoff} }
\end{figure}

For a specific parametrization of this setup, Figure \ref{fig:cutoff} shows how the optimal $\pi(A)$ and $\pi(B)$ varies as $M \in \mathbb R$ varies. 
When $M >0$, then as $\pi(A) - \pi(B)$ increases, individuals with a positive treatment effect are incentivized to report $A$ instead of $B$. This strategic behavior increases $V(\pi)$ and leads to $s(\pi, A) >0$ and $s(\pi, B) < 0$. By Corollary \ref{cor:tilde}, the cutoff rule remains optimal. When $M < 0$, however, then as $\pi(A) - \pi(B)$ increase, individuals with a negative treatment effect are incentivized to report $A$ instead of $B$. This strategic behavior decrease $V(\pi)$ and leads to $s(\pi, A)$ which can be negative, if $M$ is small enough and the proportion of strategic agents is large enough. We can see that as $M$ decreases, the optimal value for $\pi(A)$ involves randomization. As $-M$ grows very large, then $\pi(A)$ decreases towards 0. The cutoff rule is no longer optimal, since it leads to a distribution where the average agent that reports $B$ and the average agent that  reports $A$ both have a small or negative treatment effect. By reducing the probability that an agent with signal $A$ receives the treatment, randomization discourages agents with $M < 0$ from choosing the costly signal $A$ instead of the cheaper signal $B$. This incentivizes the negative treatment effect agents to continue to report $B$, which separates the negative and positive treatment effect individuals by observed signal, leading to a joint distribution of $X_i^{\pi}$ and $\tau_i$ that is more informative. 

In Section \ref{sec:numeric}, we introduce an example of a coupon targeting model where a randomization rule is optimal and a product upgrade model where the cutoff rule remains optimal.  In general, a planner needs to expand his search space beyond cutoff rules to find the optimal treatment allocation.

\subsection{Comparison to Distributionally-Robust Targeting} 
Before we introduce the numerical examples of Section \ref{sec:numeric}, we describe briefly the connection of the Stackelberg optimal rule to the min-max optimal rules that appear in the distributionally robust policy optimization literature \citet{mo2021learning, kallus2021minimax, kallus2022doubly, adjaho2022externally}. Strategic behavior results in a shift in the joint distribution in $X_i$ and $\tau_i$, where $\tau_i = Y_i(1) - Y_i(0)$. Under strategic behavior, we can define the joint distribution of the observed signal and the individual treatment effect as $F^{\pi}_{x, \tau}(x, \tau)$. For each possible treatment rule $\pi \in \Omega$, then a single joint distribution is realized. We can define the class of distributions that is formed by all possible treatment rules $\pi \in \Omega$ as $\mathcal Q=  \{ F^{\pi}_{x, \tau}: \pi \in \Omega \}$. The strategic approach chooses the treatment rule that maximizes expected outcomes, recognizing that each treatment rule is associated with a potentially distinct element of $\mathcal Q$: 
\[ \pi^* = \arg \max_{\pi} \mathbb E_{F^{\pi}_{x, \tau}} [Y_i(W_i)]. \]

The distributionally robust approach, in contrast, observes a joint distribution of $X_i$ and $\tau_i$. Then, the optimal policy maximizes the minimum expected outcome over all of the joint distributions of covariates and potential outcomes that are nearby the observed distribution, in some sense.  For example, \citet{adjaho2022externally} computes the optimal rule over all distributions of $\{ X_i, Y_i(1), Y_i(0) \} $ that have a Wasserstein distance that is less than $\epsilon$ from the observed rule. We define $\mathcal H: \{ F_{x, \tau} : d(F_{x, \tau}, \hat F_{x, \tau})  \leq \epsilon \}$, where $\hat F_{x, \tau}$ is the observed data distribution. 
\[ \pi^{DR} = \arg \max_{\pi} \min_{F \in \mathcal H} \mathbb E_{F}[Y_i(W_i)]. \] 

This is a different objective that in general will lead to a distinct optimal rule. In exploring the connection to the distributionally robust literature, however, one rule that is interesting to analyze further is the rule that maximizes the minimum expected outcomes over all of the joint distributions that can be generated by strategic behavior in response to $\pi \in \Omega$. Rather than optimizing outcomes when each treatment rule is associated with a single joint distribution of $X_i^{\pi}$ and $\tau_i$, this treatment rule is robust to strategic behavior that can occur in response to any $\pi \in \Omega$: 
\[ \pi^{R} = \arg \max_{\pi} \min_{F \in \mathcal Q} \mathbb E_{F}[Y_i(W_i)], \]
where $\mathcal Q = \{ F^{\pi}_{x, \tau} : \pi \in \Omega \}$. 

\begin{proposition} \label{prop:dro} 
Assume that $X_i \in \mathcal X$ is discrete. Assume for all $x \in \mathcal X$ and for all $\pi, \pi' \in \Omega \times \Omega$ that $\mbox{sgn}(\tau_x(\pi', x)) = \mbox{sgn}(\tau_x(\pi, x))$. Then, the min-max optimal rule over $\mathcal Q$ has a cutoff structure: 
\[ \pi^R = \mathbbm{1}(\tau_x(\pi, x) > 0) \] 
\end{proposition} 

The proof is in Appendix \ref{pf:dro}. Proposition \ref{prop:dro} indicates that in settings where strategic behavior results in changes to the magnitude of the CATEs but not their signs, then the min-max optimal rule over all joint distributions of $X_i^{\pi}$ and $\tau_i$ induced by strategic behavior has a cutoff structure. Corollary \ref{cor:tilde} indicates that the Stackelberg optimal rule, which optimizes outcomes in a framework where each treatment rule is associated with a single joint distribution of $X_i^{\pi}$ and $\tau_i$, rather than a set of distributions, can have a randomization structure and lead to a higher objective value than the min-max cutoff rule. This is demonstrated numerically in Section \ref{ex:coupon}. In settings where strategic behavior of agents is different in the deployment setting compared to the measurement setting, however, incorporating a min-max objective into the Stackelberg model may lead to a more robust policy. We leave this exploration for future work.

\section{Numerical Examples} 
\label{sec:numeric} 
We illustrate the theoretical results of the previous Section through some simple models of practical targeting problems. The first two have a single binary covariate and follow Assumption \ref{ass:strat}. The first is a model of coupon allocation and product demand where the optimal rule involves randomization in the allocation. The second is a model of a product upgrade offer where where the optimal rule takes the form of a cutoff rule despite the presence of strategic behavior. The third extends the first model to include continuous covariates. In this section, a uniform allocation policy is one where all individuals receive the treatment with the same probability. A cutoff rule is one that assigns individuals to treatment either with probability 1 or zero. The optimal rule is one that maximizes expected outcomes.

\subsection{Price Discrimination through Coupons} 

\label{ex:coupon}
In the first example, which follows Assumption \ref{ass:strat}, a profit-maximizing firm is offering a $50\%$ off coupon to certain customers for a product that normally has a per-unit profit of $\$10$. 

 For each customer $i \in [n]$, there is an unobserved indicator $\gamma_i \sim \mbox{Bernoulli}(0.50)$ that determines the distribution of potential outcomes and strategic behavior. Customers with $\gamma_i =0$ are always buyers, who will buy the product with or without the coupon. Customers with $\gamma_i = 1$ are reluctant buyers, who will not purchase the product without the coupon but will purchase it with 75\% probability if they receive a coupon. The firm has profit of \$10 for the product without the coupon and \$5 with the coupon. Since the firm is profit-maximizing, we can define outcomes
\[ Y_i(W_i) = 3.75 \cdot \gamma_i W_i + 5 \cdot ( 1- \gamma_i) W_i+ 10 \cdot ( 1 - \gamma_i) (1 - W_i). \] 

Customers with $\gamma_i = 0$ have an individual treatment effect $\tau_i = -5$ and customers with $\gamma_i = 1$ have $\tau_i = 3.75$. Rather than $\tau_i$, the store observes $X_i \in \{B, A \}$, where $X_i =A$ indicates the customer has left a product in the cart for more than a few minutes. In the absence of incentives otherwise, always buyers will purchase the item immediately, so they will have $X_i = B$. No matter the allocation policy, reluctant buyers will always leave the product in their cart and report $X_i = A$. However, when coupons are allocated using $X_i$, this introduces incentives for always-buyers to mimic reluctant buyers by leaving the product in their cart to obtain a coupon and purchase the product  for \$5 less. We model these incentives using utility function $u_1(x, \pi, \theta_i) = \pi(x) V_i - C_{ix} $ from Proposition \ref{prop:utility}.  The value of the treatment is $\$5$ when an individual buys the product. We assume that reluctant buyers, those with $\gamma_i = 1$, have $ C_{iB} > V_i$ and $C_{iA} = 0$, so they always report $X_i = A$. Always-buyers, those with $\gamma_i = 0$, have $C_{iB} = 0$, and $C_{iA} \sim \mbox{Uniform}(0, 10)$. 

This utility function leads to a signal reporting function 
\[ x(\pi, \theta_i) = B + \Big(  \mathbbm{1}(5(\pi(A) - \pi(B)) \geq C_{iA} - C_{iB} )(A - B)\Big), \]

where the individual's unobserved type is $\theta_i = \{Y_i(1), Y_i(0), C_{iB}, C_{iA}, V_i \}$ depends on $\gamma_i$. Since this model follows Assumption \ref{ass:strat} with $\rho_L = 0$, $V_M = 5$, and $\rho_M = \rho_H = 0.5$, then we can use Appendix \ref{pf:binary} to verify that the results of Theorem \ref{thm:first} apply. 

The marginal distribution of $X_i$ conditional on $\tau_i = -5$ now depends on $\pi(A) - \pi(B)$: 
\[ Pr(X_i(\pi) = A | \tau_i= -5) = Pr\Big(C_{iA} \leq 5(\pi(A) - \pi(B))\Big) \] 
Since $C_{iA}$ is uniformly distributed, then the derivative of the conditional marginal distribution exists and depends on the uniform density function. For $\pi(A)$, 
\[ \frac{\partial Pr(X_i(\pi) = A | \tau_i= -5) } { \partial \pi(A) } = \frac{1}{2} \mathbbm{1} (\pi(A) \geq \pi(B))\] 
 
The optimal policy is the coupon allocation procedure that maximizes profit when treatments are allocated on the basis of $\pi(A)$ and $\pi(B)$ and the distribution of $X_i$ is determined by $X_i(\pi)$:
\[ \pi^* = \arg \max \limits_{\pi} \mathbb E[Y_i(W_i)]. \]

Table \ref{tab:coupon} describes the performance of four different allocation rules in this model of coupon allocation. Under a uniform treatment assignment policy, $\tau_x(\pi^0, A) = 3.75 $, $\tau_x(\pi^0, B) = -5$, and the objective value is \$4.688. A cutoff rule $\pi^c(A) = 1$ and $\pi^c(B)  = 0$ raises the expected outcomes to \$5.626.  However, since $\pi^c(A) \neq \pi^c(B)$, strategic behavior is now induced in those with $\gamma_i = 0$. We derive the optimal rule of $\pi^*(H) = 0.75$ and $\pi^*(L) = 0$, which meets the condition of Theorem \ref{thm:first} and leads to higher expected outcomes than the cutoff rule. Furthermore, the Stackelberg optimal rule, which recognizes that each treatment rule is associated with a different distribution of $X_i^{\pi}$ and $\tau_i$, improves outcomes compared to a min-max optimal rule, which maximizes the minimum outcomes over $\mathcal Q = \{ F^{\pi}_{x, \tau} : \pi \in \Omega \}$. 

In this model, individuals with a negative individual treatment effect are strategic, and the incentives of the planner and agents are not aligned. As a result of this, the optimal Stackelberg rule involves some randomization in allocation; this reduces the amount of strategic behavior that occurs and ensures enough information about an individuals treatment effect is retained from the binary signal $X_i$. 

We can consider as the objective function $V(\pi)$ as a measure of the welfare of the planner. Depending on how correlated the outcome $Y_i(W_i)$ is with agent utility, raising the planner welfare may not necessarily raise agent utility. The last row of the table reports $\mathbb E[u_1(x(\pi, \theta_i), \pi, \theta_i)]$ as a measure of consumer welfare for each allocation rule. The optimal Stackelberg rule raises the planner's objective function but lowers agent utility compared to the uniform rule that doesn't distinguish between agents. This is not always the case, however. In our next example, the optimal Stackelberg rule raises both the planner's objective function and agent utility compared to the uniform rule. 

\begin{table}
\centering
\begin{tabular} { | l |  r r r r  | } 
\hline
& Uniform ($\pi^0$) & Cutoff  ($\pi^c$) & Min-Max ($\pi^R$) & Optimal Stackelberg ($\pi^*$) \\ 
\hline 
$\pi(B)$ & 0.5 & 0.0 & 0.0 & 0.0 \\ 
$\pi(A)$ & 0.5 & 1.0 & 1.0 & 0.75 \\
\hline
$\tau_x(\pi, B)$& -5.00 & -5.00 & -5.00 & -5.00 \\ 
$\tau_x(\pi, A)$ & 3.75 & 0.83& 0.83 & 1.36  \\ 
\hline 
$V(\pi)$  & \textbf{4.688} & \textbf{5.625} & \textbf{5.625}& \textbf{5.703} \\
$\mathbb E[u_1(\theta_i)]$ & 2.1875 & 2.50 & 2.50  & 1.875  \\ 
\hline   
\end{tabular} 
\caption{ Performance of Allocation Rules in Coupon Model\label{tab:coupon} }
\end{table}

\subsection{Allocating Product Upgrades} 
\label{ex:upgrade}

In this second example, a firm is offering a product upgrade for purchase. Again, there are $n$ customers and for each customer $i \in [n]$, $\gamma_i \sim \mbox{Bernoulli}(0.5)$ determines potential outcomes and the distribution of strategic behavior.  $\gamma_i =0$ indicates basic customers and $\gamma_i = 1$ indicates advanced customers. $W_i = 1$ indicates that customer $i$ receives an offer of a product upgrade. The firm receives a profit of \$5 from each customer who is not treated. The upgrade offer annoys basic customers who have no use for the upgrade, so they reduce their usage upon receiving treatment, leading to a profit of \$1 per basic customer who is treated. For advanced customers, they benefit from the product upgrade and purchase it, leading to a profit of \$10 per advanced customer who is treated: 

\[Y_i(W_i) = 5 (1-W_i) + (1- W_i) (10 \gamma_i - (1 - \gamma_i)).\] 
$\tau_i = -4$ for basic customers and $\tau_i = 5$ for advanced customers. The firm does not observe customer sophistication directly, but they do observe whether or not they have completed a certification course $X_i \in \{ B, A \}$. Basic customers find it costly to complete the certification course and do not have value for the product upgrade, so they always report $X_i = B$. Advanced customers, even in the absence of outside incentives, sometimes will complete the course anyway. They also have value of \$5 for the product upgrade, so when the treatment is targeted based on course completion, this incentivizes additional advanced customers to complete the course. 

We again model these incentives using utility function $u_1(x, \pi, \theta_i) = \pi(x) V_i - C_{ix} $ from Proposition \ref{prop:utility}.  $C_{iB} = 0$ for all customers. The basic customers have $V_i = 0$ and $C_{iA} > 0$. For advanced buyers, $C_{iA} \sim \mbox{Uniform}(-10, 10)$ and $V_i =5$. As in the previous example, 
\[ x(\pi, \theta_i) = B + (A - B) \mathbbm{1}(5(\pi(A) - \pi(B)) > C_{iA}). \] 

Since this model follows Assumption \ref{ass:strat} with $\rho_H = 0$, $\rho_M = \rho_L = 0.5$, and $V_M =5$, then we can use Appendix \ref{pf:binary} to verify that the results of Theorem \ref{thm:first} apply. The optimal policy is the upgrade allocation procedure that maximizes profit when treatments are allocated on the basis of $\pi(A)$ and $\pi(B)$ and the distribution of $X_i$ is determined by $X_i(\pi)$:
\[ \pi^* = \arg \max \limits_{\pi} \mathbb E[Y_i(W_i)]. \] 

In this example, those with a positive treatment effect are strategic. As the planner targets the offer more precisely to those who complete the certification course, they incentivize those who would benefit the most from the offer to complete the certification course. As a result, as we expect from Proposition \ref{prop:suf}, the optimal Stackelberg rule in this model is a cutoff rule in the presence of beneficial strategic behavior; see Table \ref{tab:upgrade} for a summary and Appendix \ref{app:upgrade} for a derivation of the optimum and confirmation that this example meets the differentiability conditions of Proposition \ref{prop:suf}. In this example,  the optimal Stackelberg rule improves  customer welfare compared to uniform targeting, since advanced customers, who have utility for the product upgrade offer, receive the offer more often.

\begin{table}
\centering
\begin{tabular} { | l |  r r r  | } 
\hline
& Uniform Rule($\pi^0$) & Cutoff Rule ($\pi^c$) & Optimal Stackelberg Rule ($\pi^*$) \\ 
\hline 
$\pi(B)$ & 0.5 & 0.0 & 0.0 \\ 
$\pi(A)$ & 0.5 & 1.0 & 1.0 \\
\hline
$\tau_x( B, \pi)$& -1.00 & -1.20 & -1.20 \\ 
$\tau_x(A, \pi)$ & 5.00 & 5.00 & 5.00  \\ 
\hline 
$V(\pi)$  & \textbf{5.25} & \textbf{6.875} & \textbf{6.875} \\
$\mathbb E[u_1(\theta_i)]$ & 1.875 & 2.812 & 2.812  \\
\hline   
\end{tabular} 
\caption{ Performance of Allocation Rules in Product Upgrade Model \label{tab:upgrade} }
\end{table}

\subsection{Coupon Allocation with Continuous Covariates} 

In the previous two subsections, we illustrated our theoretical results in a simple model where there is a single binary covariate. A natural question is how these results extend to a setting with continuous covariates, with $X_i \in \mathbb R$. For this numerical example, so that the planner's optimization problem remains finite dimensional, we restrict the treatment allocation rule to be a parametric function. In particular, we choose a logit function here, so that 
\[ \pi(X_i; \bm \beta) = \frac{1}{1 + e^{- (\beta_0 + X_i \beta_1)}}. \]

The goal is to find the $\beta$ that maximizes expected outcomes: 
\[ \bm \beta^* = \arg \max \limits_{\bm \beta \in \mathbb R^2} \mathbb E [Y_i(W_i] \] 

For this section, we construct a modified version of the coupon allocation model. As in the discrete setting, agents have unobserved discrete types $\gamma_i \in \{0, 1\}$, where $\gamma_i = 0$ represents always-buyers with a negative ITE and $\gamma_i = 1$ represents reluctant buyers with a positive ITE. $\theta_i$ is made up of $Z_i$ and $C_i$. $Z_i \in \mathbb R$ representing the agent's inherent, costless behavior, where $Z_i \sim \mbox{Normal}(-1, 2)$ if $\gamma_i = 0$ and $Z_i \sim \mbox{Normal}(1, 2)$ if $\gamma_i = 1$.  Always-buyers, who are potentially strategic, have utility functions
\[ u(x, \pi, \theta_i) = 5\cdot \pi(x; \beta) -  C_i (x - Z_i)^2, \]
while reluctant buyers always report $X_i = Z_i$. The utility is made up of a value for receiving the treatment and a cost of reporting covariate $x$. Following the literature on prediction with strategic behavior, we have assumed in our definition of the utility function that the individual has some inherent behavior $Z_i$ and that the cost for reporting a different covariate $x$ is quadratic in the distance from $Z_i$ \citep{frankel2019improving}. This utility specification leads to a reporting rule 
\[ X_i(\pi) = \gamma_i Z_i + ( 1-  \gamma_i) \arg \max \limits_{x \in \mathbb R} U_i(x, \pi). \]  
As in the discrete setting, $C_i \in \mbox{Uniform}(0, 10)$ and potential outcomes are
\[ Y_i(W_i) = 5 \cdot (0.75 \gamma_i + ( 1- \gamma_i)) W_i+ 10 \cdot ( 1 - \gamma_i) (1 - W_i). \] 

\begin{figure}[ht]
\centering
\begin{subfigure}[b]{0.49\textwidth}
         \includegraphics[width=\textwidth]{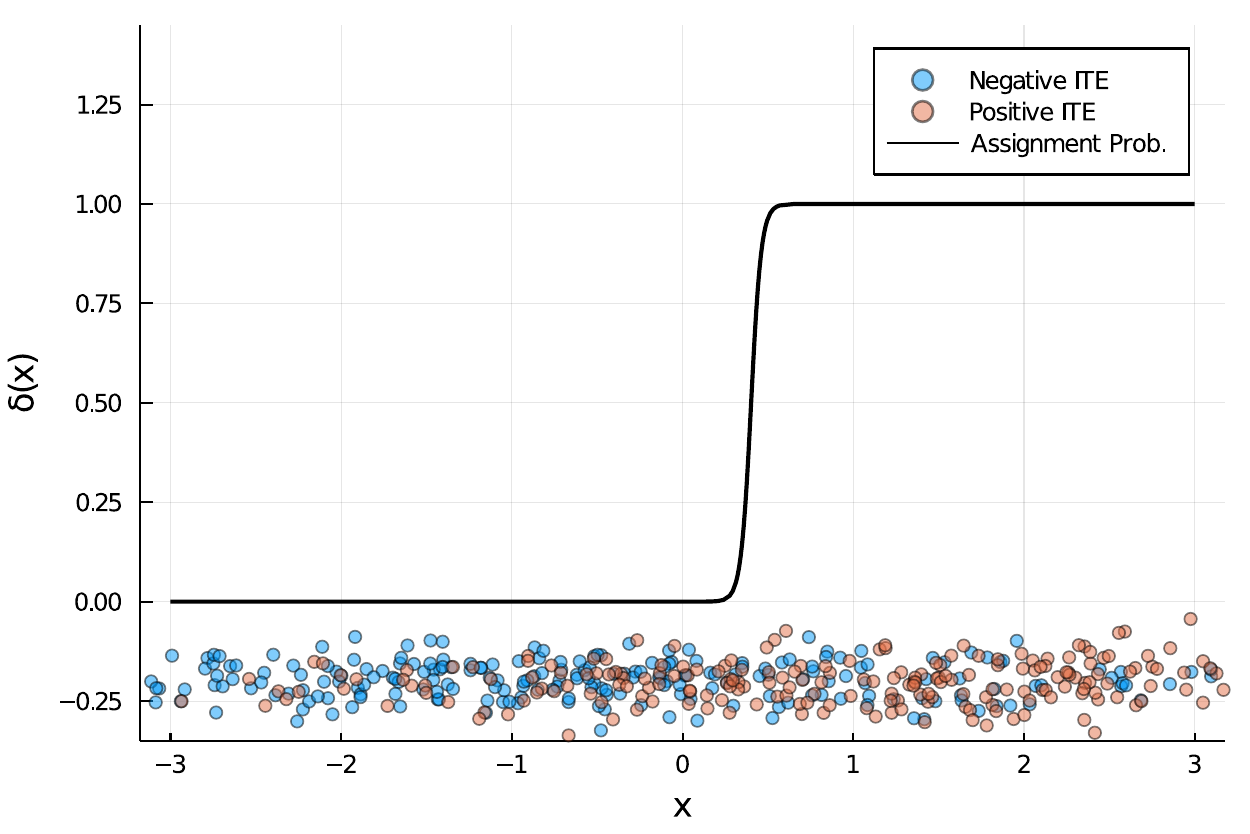}
         \caption{Cutoff Rule without Strategic Behavior}
         \label{fig:nonstrat}
     \end{subfigure}
     \begin{subfigure}[b]{0.49\textwidth}
         \centering
         \includegraphics[width=\textwidth]{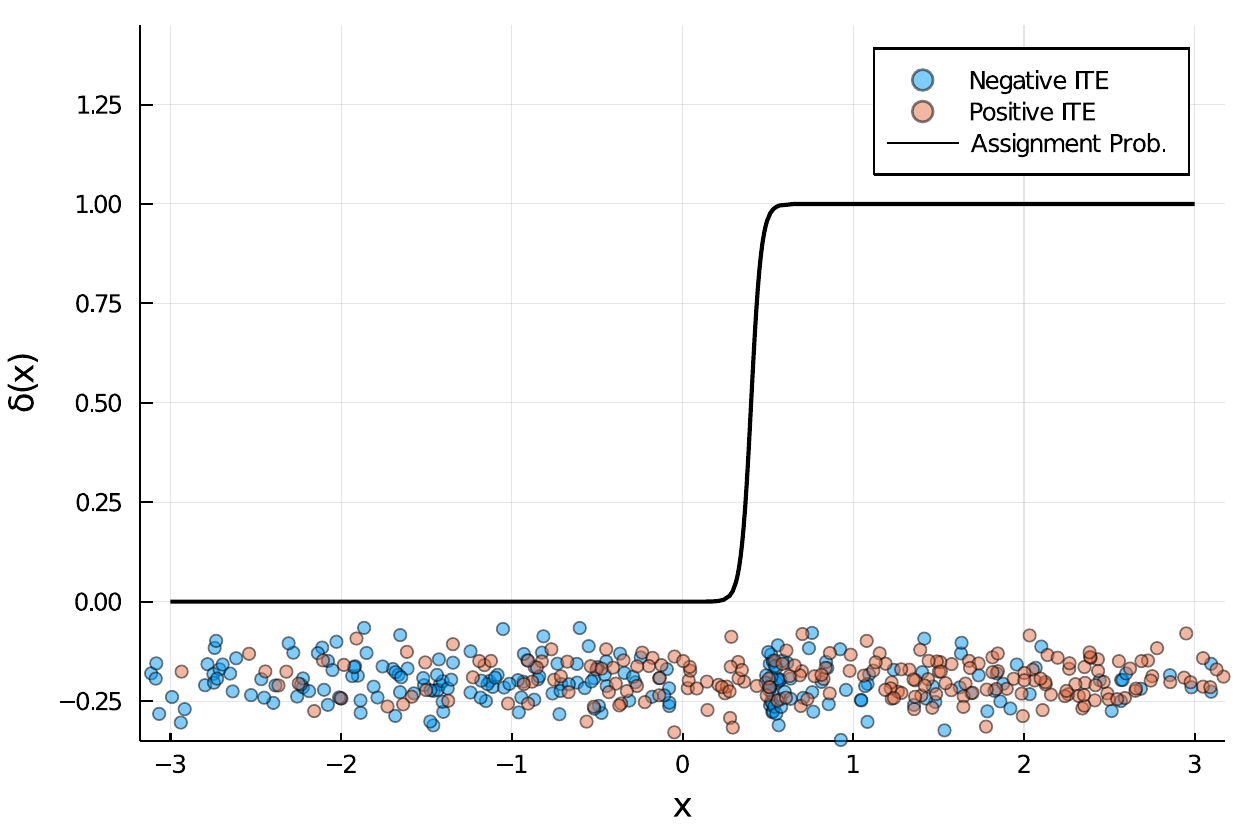}
         \caption{Cutoff Rule with Strategic Behavior}
         \label{fig:stratces}
     \end{subfigure}
     \begin{subfigure}[b]{0.49\textwidth}
         \centering
         \includegraphics[width=\textwidth]{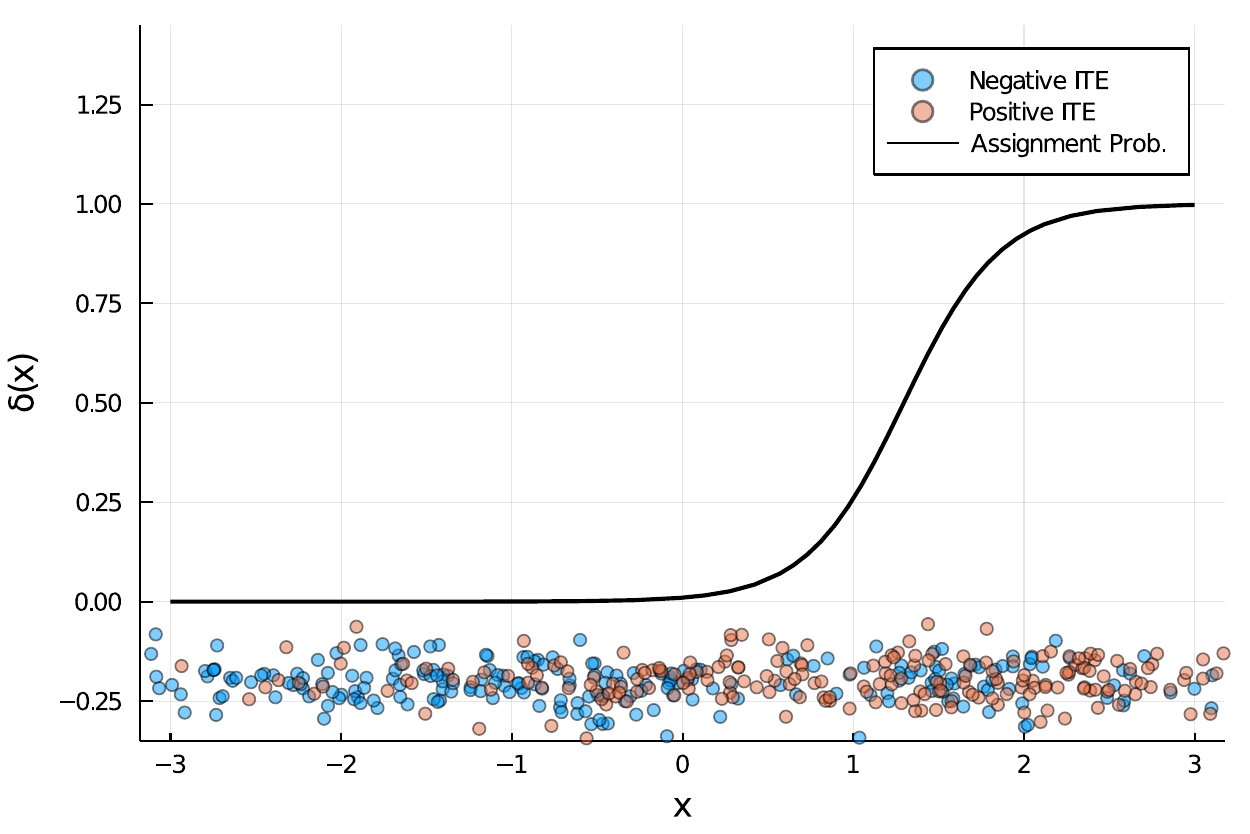}
         \caption{Optimal Stackelberg Rule}
         \label{fig:stratopt}
     \end{subfigure}
 
\caption{These figures plot the probability of being assigned a coupon conditional on reporting $X_i$ under different scenarios. The reported $X_i$ for each individual, colored by their ITE, is plotted using a jitter underneath the allocation rules.     \label{fig:contmodel} } 
\end{figure}

Figure \ref{fig:contmodel} plots the allocation rule and the resulting response of individuals for three different settings. Underneath the graph of the treatment rule, the locations of 200 individuals who respond to this treatment rule are plotted with a vertical jitter, colored by their ITEs. For Figure \ref{fig:nonstrat}, we simulate a version of the coupon model with continuous covariates and without strategic behavior, so that $X_i = Z_i$ and those with a lower $X_i$ are more likely to have a negative ITE. We plot the resulting optimal rule, and as expected, the optimal logit function is close to a step function, and the step occurs where the CATE transitions from negative to positive. The average profit per agent is \$5.72. If this cutoff function is implemented in a population who report $X_i$ strategically based on the model in this section, as in Figure \ref{fig:stratces}, then the distribution of $X_i$ shifts. Those with $\gamma_i = 0$ but with inherent behavior $Z_i$ that is close to the cutoff will shift their behavior to report $X_i >0$ and receive the valuable coupon. As a result, the profit drops to \$5.49 per agent. Taking into account the strategic behavior in Figure \ref{fig:stratopt}, the optimal logit function is no longer a cutoff rule. Instead, there is a fuzzy region where treatments are assigned with below one probability to those with positive $X_i$. There is still some strategic behavior induced (some agents with a negative ITE report $X_i > Z_i$), but the average profit of $\$5.55$ per agent is improved. 

Just as in the binary case, with continuous covariates, a rule that assigns treatment with 100\% probability to groups with a positive CATE may induce a distribution where targeting is far less effective than one that commits to treating individuals in a more uniform way. Knowing that the optimal Stackelberg rule may not take the form of a cutoff rule in both continuous and discrete settings, our next goal is coming up with estimation methods for  $\pi^*$. In the next section, we show how estimating the optimal rule can be described as a type of sequential optimization problem with noisy function evaluations, and we design  a procedure that recovers the optimal rule without  relying on strong assumptions on the structure of strategic behavior. 

\section{Estimating the Optimal Treatment Rule} 
\label{sec:est} 

In order to understand how strategic behavior incentivized by changes in the treatment rule impacts the objective, it is helpful to have some variation in the treatment rule. To estimate the rule in finite samples without a model for strategic behavior, we assume that the planner is in a sequential experimentation setting. 

\subsection{Sequential Experimentation} 

\begin{assumption} \textbf{Optimization Environment: } $X_i \in \mathcal X$ is discrete, with $| \mathcal X | = d$. At each time $t = 1, \ldots, T$, a batch of $n$ agents arrive and the Stackelberg game from Section \ref{sec:model} is played for each $ i \in [n]$: 
\begin{enumerate} 
\item Agents report signal $X^t_i = x(\pi^t, \theta_i)$ 
\item Treatments are allocated, with $Pr(W^t_i = 1) = \pi^t(X^t_{i})$ 
\item Outcomes are observed, with $Y^t_i = Y^t_i(W^t_i)$. 
\end{enumerate} 
\label{ass:env} 
\end{assumption}

We restrict the treatment rule to be constant within each time step $t$, so we do not consider procedures that randomize different allocation rules to different agents within the same time period. Although we don't model this explicitly, this allows agents to take some time to collectively recognize and respond (in possibly heterogeneous ways) to $\pi^t$ before outcomes are measured. If learning occurs, it must be complete before the outcomes are evaluated at step $t$. This kind of collective learning at each time step is much more challenging if each agent was assigned a different allocation rule. One downside of this restriction, however, is that if there is time-varying noise in the environment with meaningful drift or seasonality, then evaluating only one policy at each time step can make optimization challenging. One way that multiple policies could be evaluated simultaneously is if agents were grouped in clusters, where collective learning about a treatment rule occurred within, but not across clusters. Treatment rules could then be assigned at a cluster level, which allows more than one policy to be evaluated in a single time step, as in \citet{letham2019constrained}. For the purposes of this paper, we focus on the sequential optimization approach and leave the batch optimization framework and a more in-depth analysis of agent learning for future work. 

At each time $t = 1, \ldots, T$, a group of of $n$ agents arrive and are treated by the planner.  In the coupon example, we can think of  the group of agents that arrive at each time $t$ as the customers who arrive to the seller's website within a fixed time period. We assume that the agent's decision problem is static, so we can ignore dynamic considerations if an agent arrives repeatedly over time. Through the heterogeneity in choice functions $x(\pi^t, \theta_i)$, our framework allows for heterogeneity in strategic behavior by agents, including in how accurately they respond to $\pi^t$. 

In this environment, the planner cannot observe $V(\pi)$ directly when there are a finite sample of agents at each time step. However, they can observe a noisy version of $V(\pi)$, 
\[ V_n(\pi^t) = \frac{1}{n} \sum \limits_{i=1}^n Y^t_i(W^t_i).  \]
As $n$ grows large, we have from the Central Limit Theorem that 
\[ \sqrt n( V_n(\pi^t) - V(\pi^t)) \rightarrow_D N(0, \sigma^2). \] 
At time-step $t$, a large sample approximation of our setting is that the planner observes $V_n(\pi^t) = V(\pi^t) + \epsilon$ where $\epsilon$ is normally distributed. The goal is a procedure for setting $\pi^t$ that makes cumulative regret small, defined as
\[ R_T = \sum \limits_{t=1}^T [ V(\pi^*)  - V( \pi^t)], \] 
 where $\pi^* \in \arg \max \limits_{\pi \in [0, 1]^d} V( \pi)$. 

This is a standard setting of sequential optimization where only noisy function evaluations are available. Our choice of an optimization method for $V(\pi)$ is guided by a few factors. Without any continuity assumptions on $V(\pi)$, this optimization problem is NP-Hard. With a strong-concavity assumption on the objective, we can use an approximate gradient approach based on function evaluations, as in \citet{flaxman2004online}. However, it is challenging to verify concavity assumptions on the objective $V(\pi)$, since it depends on the unknown structure of strategic behavior. As a result, we make smoothness assumptions but do not make concavity assumptions, and thus do not use a gradient approximation-based approach. Second, function evaluation is costly. Although we do not model this cost explicitly, each time a platform changes its targeting rule, it requires engineering effort and time for customers to adjust to the new rule. As a result, we would prefer a method that finds an optimal rule with a limited number of time steps. 

Bayesian Optimization has been used successfully to find hyperparameters that optimize unknown non-concave functions that are costly to evaluate in machine learning settings \citep{snoek2012practical}. \citet{letham2019constrained}  uses Bayesian Optimization to optimize a complex class of system configuration parameters by running batches of A/B tests. Our preferred approach for optimizing targeting rules when agents are strategic is also based on Bayesian Optimization, and is briefly described in the next section.

\subsection{Bayesian Optimization} 

Bayesian Optimization uses a Bayesian decision-making framework to optimize an unknown function using sequential evaluations. There are two choices that must be made to implement Bayesian optimization for a specific problem. The first is specifying a prior that defines a probability distribution for the unknown objective function. The prior should have non-zero support over a function space that is limited in some ways (e.g. to functions of a certain smoothness), but flexible enough that it contains the true objective function. In Bayesian optimization, the posterior distribution for the objective is updated as the objective is evaluated sequentially. Choosing where to evaluate the objective function given a posterior distribution for the objective relies on a choice of acquisition function. An acquisition function maps the posterior distribution to a utility value for each possible evaluation point. At a given step in the optimization procedure, the next evaluation point is chosen by maximizing the acquisition function. 

We choose a Gaussian Process prior, for their flexibility and tractability. Formally, our unknown objective function $V(\pi)$, where $\pi \in [0, 1]^d$, is drawn from a Gaussian Process \citep{williams2006gaussian}, defined by a mean function $\mu: [0, 1]^d \rightarrow \mathbb R$ and kernel function $k: [0, 1]^d \times [0, 1]^d \rightarrow \mathbb R$. The choice of kernel function enforces certain restrictions on $V$; for example if the kernel used is the squared exponential kernel, then the objective function is assumed to be in class of infinitely differentiable functions $C^{\infty}$.  If a function $f$ evaluated with independent normal noise with variance $\sigma^2$ is given a prior distribution given by $GP( \mu_0, k)$, then the prior distribution of a vector of function values $\bm Z = f(\bm Q)$ is multivariate normal, so that $\bm Z | \bm Q \sim N(\mu_0(\bm Q), k(\bm Q, \bm Q) + \sigma^2 \bm I)$.  Given a set of noisy function evaluations, then the posterior distribution of the function also follows a Gaussian Process. 

Let $\pi^s = [ V_n(\pi_t)]_{t=1}^s$ be a vector of $s$ sequential noisy function evaluations, where $\pi^s_t = V_n(\pi_t)$. $\bm D^s = [ \pi_t ]_{t=1}^s$ is the matrix defining $s$ allocation functions, where $D^s_{tk} = Pr(W^t_i = 1 | X^t_i = k)$.  Given a set of noisy function evaluations, the posterior distribution for the unknown objective function is a Gaussian Process with mean function $\mu_s$ and kernel function $k_s$. 
\[ \mu_s(\pi) = \mu_0(\pi) + k(\pi, \bm D^s) ( \Sigma + \sigma^2 \bm I) (\pi^s - \mu(\bm D^s)), \] 
\[ k_s(\pi, \pi') = k(\pi, \pi') - k(\pi, \bm D^s) ( \bm \Sigma^s + \sigma^2 \bm I)k(\bm D^s, \pi), \]

where $\bm \Sigma^s = k( \bm D^s, \bm D^s)$ is the $s \times s$ kernel matrix. 

The next step is to choose an acquisition function, which determines which point to evaluate at time $t$ given a Gaussian Process posterior constructed from a prior and observations up to time $t-1$. There are a variety of acquisition functions proposed in the literature, see \citet{garnett2023bayesian} unified treatment. Popular choices include Expected Improvement, which chooses the point that has the largest expected improvement in the objective over the current best point, Probability of Improvement, which chooses the point that maximizes the probability of improving over the current best function value, and Upper Confidence Bound approaches. For the purposes of the simulations in Section \ref{sec:mturk}, we choose the upper confidence bound (UCB) algorithm since it takes a tractable and intuitive form even with observation noise, and the UCB approach comes with formal regret guarantees \citep{srinivas2009gaussian}. Given the wide range of practical problems which have been successfully addressed using expected improvement, see \citet{snoek2012practical} and \citet{letham2019constrained}, we expect that an approach based on expected improvement would also work well in our setting. 

With a choice of Gaussian Process prior, and the UCB acquisition function, Bayesian optimization takes the form of Algorithm 1 \citep{auer2002using, dani2008stochastic, srinivas2009gaussian}. 

\begin{algorithm}[!ht] 
\KwIn{Prior $GP(\mu_0, k)$} 
\KwOut{Estimate of treatment rule $\pi_T$} 
\For{$t \in \{1, \ldots T \}$} 
{ 
	Choose $\pi_t = \arg \max \limits_{\pi \in [0,1]^d} \mu_{t-1}(\pi) + \sqrt \alpha_t \sqrt {k_{t-1}(\pi,  \pi)}$; \\
	Receive feedback $\pi_t = V_n(\pi_t)$;  \\
	Compute $\mu_t$ and $k_t$ based on Bayesian update;  
} 
\caption{Bayesian Optimization with Gaussian Processes \label{alg:ucb}} 
\end{algorithm}

This algorithm favors points that have a high posterior mean, where it appears the function has a maximum, and a high posterior variance, where there is substantial uncertainty about the function value. $\alpha_t$ is a tuning parameter which determines the exploration-exploitation tradeoff. Under a selection procedure for $\alpha_t$, \citet{srinivas2009gaussian} provides bounds on the cumulative regret of the GP-UCB algorithm for a class of kernel functions. In the next section, we use the squared exponential kernel in our prior; if it is the case that the objective $V(\pi)$ has a positive probability under this prior, then the results of \citet{srinivas2009gaussian} dictate that regret grows at a sublinear rate, i.e. $R_T = O\left(\sqrt{T(\log T)^{d+1}} \right)$. 

Algorithm \ref{alg:ucb} represents a sequential experiment that can be used to optimize a targeting rule over time. The planner starts with a prior on how the objective function varies with respect to the parameters of the targeting rule. The targeting rule is varied over time based on optimization an acquisition function, the posterior for the objective function is updated using the average outcomes of individuals observed at a given time step, and uncertainty about where the global maximum of the function is located is reduced. The algorithm eventually converges to the globally optimal rule, as long as the objective function has the required smoothness. Concavity of the objective function, which may not always hold under strategic behavior, is not required for the regret properties to hold. We next illustrate the regret properties of this procedure in a semi-synthetic simulation. 

\section{MTurk Experiment}
\label{sec:mturk} 

In this section, we first describe an MTurk experiment that tests how allocating a bonus payment shifts the joint distribution of two questions in the survey. We then use the variation from the experiment to estimate a simple structural model of strategic manipulation and run a simulation that is used to evaluate the regret of Algorithm \ref{alg:ucb}. 


The experiment consisted of a survey with four questions. IRB approval was received for this experiment.\footnote{The protocol, IRB-56869, qualified for an Exempt Review.} The first asks individuals how much they like math, on a scale of 1-5. The fourth question is optional and asks the respondent to report either of the solutions to $2x^2-5x = 3$. The correct answers are -0.5 or 3. In Wave 1 of the survey, the respondents individuals are paid $\$0.10$ to complete Q1-3 of the survey. In Wave 2 of the survey, individuals are paid $\$0.10$ to complete Q1-3 of the survey, but there is a bonus payment of $\$1$ allocated if the respondent answers the fourth question correctly. The first $n=316$ respondents are analyzed for each wave of the survey. The second and third questions are not used for this empirical exercise. The language for the survey questions and the bonus payments is in Appendix \ref{app:mturk}. 

In Table \ref{tab:exp}, we report the proportion of individuals who like math and respond 4 or 5 to Question 1 for both waves of the survey. We also report proportion of individuals who respond to Q4 correctly. Last, we report the proportion of individuals who respond correctly to Q4 conditional on liking math or not. The bonus payment, which is 10 times the base payment for the survey, is a valuable treatment that is allocated conditional on the response to Q4. We denote $X_i^{\pi} = A$ if the Q4 response is correct, and $X_i^{\pi} = B$ otherwise. So, for Wave 2 of the survey, we have $\pi(A) = 1$ and $\pi(B) = 0$. For Wave 1 of the survey $\pi(A) = \pi(B) = 0$. Whether or not an individual responds to Q4 depends on how costly or rewarding they find it to work through the math problem, as well as their value for the bonus payment. We find that more individuals respond to Q4 correctly in Wave 2 than Wave 1, due to the introduction of the bonus payment.  

The response to Q1, on the other hand, does not affect the payment an individual receives for the survey, so we don't expect that there will be meaningful strategic behavior in responses to this question. As expected, the proportion of individuals who like math does not meaningfully change between Wave 1 and Wave 2.  We use the notation $\gamma_i = 1$ to indicate whether an individual likes math.  However, liking math does affect how costly an individual finds it to solve the math question, and how they respond to the bonus payment. Between Wave 1 and Wave 2, the proportion of individuals responding correctly to Q4 increases for math-dislikers by more than it does for math-likers.

\begin{table}
\centering
\begin{tabular} { | l  | l ||  r r  | } 
\hline
\textbf{Label} & \textbf{Notation} & \textbf{Wave 1} & \textbf{Wave 2} \\ 
\hline 
Bonus Probability  | Correct &  $\pi(B)$ & 0.0 & 0.0  \\ 
Bonus Probability | Incorrect & $\pi(A)$ & 0.0 & 1.0  \\
\hline
Proportion Liking Math & $Pr(\gamma_i = 1)$  & 0.43 (0.027) & 0.49 (0.027)  \\ 
Proportion Q4 Correct & $Pr(X_i = A)$ & 0.41 (0.027) & 0.60 (0.027) \\ 
Proportion Correct | Like Math & $Pr(X_i = A | \gamma_i = 1)$ & 0.56 (0.042) & 0.72 (0.036) \\ 
Proportion Correct | Don't Like Math & $Pr(X_i = A | \gamma_i = 0)$ & 0.29 (0.034) & 0.48 (0.039) \\ 
\hline   
\end{tabular} 
\caption{ Summary Statistics of MTurk Experiment. For the proportions, standard errors are in brackets. \label{tab:exp}}
\end{table} 

We now introduce a simple structural model, where the distribution of strategic behavior depends on liking math, where $\gamma_i \sim \mbox{Bernoulli}(\lambda)$.  We use the data generated from the MTurk experiment to estimate the parameters of this structural model. Then, we simulate data from the structural model to test the regret of Algorithm \ref{alg:ucb}. We assume that individuals choose whether or not to respond to Q4 correctly by maximizing their utility, where the utility function is $u_1(x, \pi, \theta_i) = V_i \pi(x) - C_{ix}$ from Proposition \ref{prop:utility}. 

We assume that all individuals value the bonus at its monetary value, so $V_i = 1$. Let $M_i \in \{0, 1\}$ determine whether an agent is strategic, where $M_i \sim \mbox{Bernoulli}(\rho_{\gamma_i})$. When $M_i = 0$, then agents enjoy solving the math question with or without the bonus, so $C_{iB} = 1$ and $C_{iA} = -1$, and $X_i^{\pi} = A$ for any $\pi$. When $M_i = 1$, then $C_{iB} = 0$ and $C_{iA} \sim \mbox{Uniform}(0, b_{\gamma_i})$. The response to Q1 is equal to $\gamma_i$, and is not affected by $\pi$. The response to Q4, on the other hand, is governed by 

\[ x(\pi, \theta_i) = B + ( A - B) (\pi(A) - \pi(B) \geq C_{iA}); \]  

We assume that the distribution of individual types who take the survey is constant across waves. Then, $\lambda$ is identified by  the observed average $Pr(\gamma_i = 1)$ across the two waves of the survey. $\rho_{\gamma_i}$ is identified by $Pr(X_i = A | \gamma_i)$ in Wave 1. $b_{\gamma_i}$ is identified by the change in $Pr(X_i =A | \gamma_i)$ across the two waves of the survey. 

Based on the moments from Table \ref{ass:strat}, the estimated parameters of the structural model are in Table \ref{tab:params}, with standard errors in brackets computed using the bootstrap. 

\begin{table} [!ht]
\centering
\begin{tabular} { | l | c c c  c c |} 
\hline
 &$\hat \lambda$ &  $\hat \rho_0$ & $ \hat \rho_1 $&$ \hat b_0$ &$ \hat b_1$ \\ 
 \hline
Estimate & 0.46 &  0.71 & 0.44 & 4.13 & 3.30 \\ 
Standard Error & (0.019) & (0.034) & (0.042) & (1.83) & (2.30)\\ 
\hline
\end{tabular}
\caption{Parameter Estimates for Semi-Synthetic Simulation \label{tab:params} }
\end{table} 

Close to half of the agents like math. Liking math impacts whether an individual responds correctly to the factoring question. Those with $\gamma_i = 1$ are much more likely to respond to Q4 correctly. Solving for the optimal treatment allocation rule in this structural model, which has a concave objective, we find that the optimal Stackelberg rule has $\pi^*(B) = 0 $ and $\pi^*(A) = 0.66$.  

Given this structural model, we would like to simulate how Algorithm \ref{alg:ucb} would perform. However, we have not yet defined outcomes. In the actual experiment, it was equally costly to allocate the bonus payment to those who like math and those who do not like math. However, we can use the structural model to simulate a setting where it was beneficial from the perspective of the planner to allocate a treatment to those who like math and it is harmful to allocate it to those who do not like math. We assume that 
\[ Y_i(W_i) = W_i (5 \gamma_i  - 6 ( 1- \gamma_i)). \] 

\begin{figure}[!ht]
 \centering
         \includegraphics[width=\textwidth]{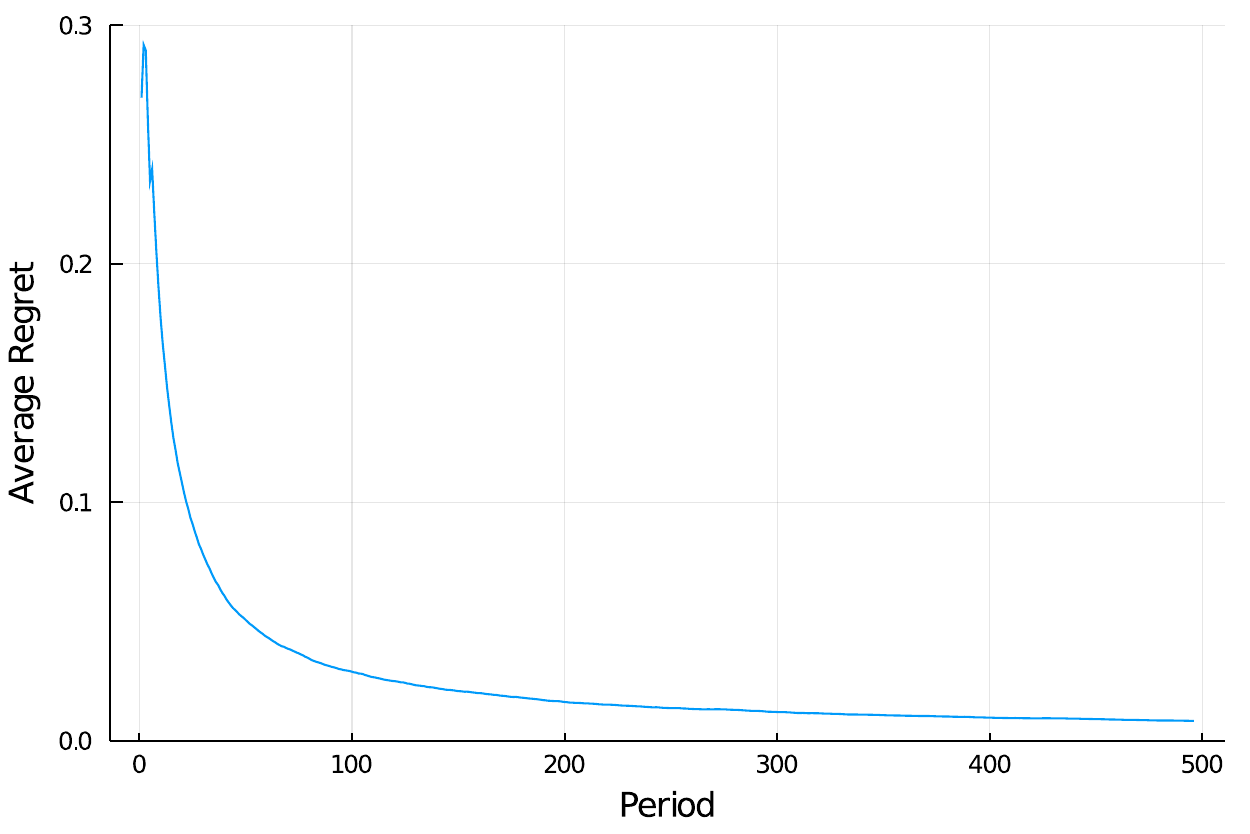}
\caption{Average Regret $R_T/T$ of Algorithm \ref{alg:ucb}   \label{fig:regb} } 
\end{figure}

This implies that those who like math have $\tau_i = 5$ and those who dislike math have $\tau_i = -6$. We assume that the algorithm is run by a planner who does not have any knowledge of the underlying structural model for strategic behavior, but can observe outcomes from a sample of $n=2000$ agents who respond according to this estimated model repeatedly over time. In Figure \ref{fig:regb}, we plot the average regret $\frac{R_T}{T}$ over 500 periods of Algorithm \ref{alg:ucb}, averaged over 50 repetitions of the simulation. 

We can now simulate how Algorithm \ref{alg:ucb} would perform. We see that Algorithm \ref{alg:ucb} quickly finds a treatment allocation rule that is close to the optimum and average regret decreases rapidly towards zero. 

\section{Conclusion}

In this paper, we show that strategic behavior impacts the structure of optimal treatment rules that allocate binary treatments based on observed covariates. Our first result indicates that it is possible that the optimal treatment rule includes randomization, in contrast to the setting without strategic behavior where a cutoff rule is always optimal. Through numerical and theoretical results, we show that the average treatment effect of certain strategic individuals determines whether or not a policy maker needs to expand their policy search space to include rules with randomization. We propose a sequential experiment based on Bayesian optimization that converges to the optimal treatment assignment function, without making parametric assumptions on the structure of individuals' strategic behavior.

Our sequential experiment captures only a limited form of dynamics, since agents have a myopic objective. An exciting avenue for future work would be understanding the impact of more complex forms of agent learning and optimization on treatment rules, see \citet{mansour2015bayesian} on coordinating agent behavior in the bandit setting. Other avenues for future work include exploring how strategic behavior impacts the analysis of causal settings with interference between units. 
 
\newpage 

\bibliography{../sample.bib}
\newpage 
\appendix

\section{Proofs}
\label{sec:proof}
\subsection{Proof of Theorem \ref{thm:first} } 
\label{pf:first} 
\begin{proof} 
The first step is to show that 
\[ \partial V(\pi ; h) = \int h(x) [f_x(\pi, x) \tau_x( \pi, x) + s(\pi, x) ]d \mu (x) \]
where $\partial V(\pi; h)$ is the Gateaux derivative of $V(\pi)$ in the direction of $h \in \Omega$. First, we write $V(\pi)$ as an integral over the joint distribution of $X_i^{\pi}$ and $\tau_i$: 
\begin{equation*}
\begin{split} 
 V(\pi)  &= \mathbb E[\pi(X^{\pi}_i) (Y_i(1) - Y_i(0))] \\ 
 & = \int \tau \cdot \pi(x) dF_{x, \tau}(x, \tau).
 \end{split} 
\end{equation*} 
We can derive the Gateaux derivative of $V(\pi)$ using the product rule: 
\begin{equation*} 
\begin{split} 
\partial V(\pi; h) & = \lim \limits_{\delta \rightarrow 0} \frac{\int \tau \cdot [ \pi(x) + \delta h(x)] dF^{\pi+ \delta h}_{x, \tau}(x, \tau)  -   \int  \tau \cdot  \pi(x)  dF^{\pi} _{x, \tau}(x, \tau) }{\delta}  \\ 
& \stackrel{(1)}{=} \lim_{\delta \rightarrow 0} \int \tau \cdot h(x)  dF^{\pi+ \delta h}_{x, \tau}(x, \tau)  +  \lim \limits_{ \delta \rightarrow 0 } \int  \tau \cdot  \pi(x)  \frac{dF^{\pi + \delta h} _{x, \tau}(x, \tau)  - dF^{\pi}_{x, \tau} (x, \tau ) }{\delta} \\ 
& \stackrel{(2)}{=} \int \tau_x(\pi, x) h(x) f_x(\pi, x) d \mu(x) +   \lim \limits_{ \delta \rightarrow 0 } \int  \tau \cdot  \pi(x)  \frac{dF^{\pi + \delta h} _{x, \tau}(x, \tau)  - dF^{\pi}_{x, \tau} (x, \tau ) }{\delta} \\ 
& \stackrel{(3)}{=} \int h(x) (f_x(\pi, x) \tau_x(\pi, x) + s(\pi, x) ) d \mu(x) 
\end{split}
\end{equation*} 

Step (1) is from Condition \ref{eqn:first}. Step (2) is from the law of iterated expectations: $\mathbb E[h(X_i^{\pi}) Y_i(1) - Y_i(0)] = \mathbb E_x[h(X_i^{\pi}) \mathbb E[Y_i(1) - Y_i(0) | X_i^{\pi}]]$. Step (3) is also from Condition \ref{eqn:first}. 

Since the vector space $\Omega$ is convex, Theorem 2 of Chapter 7 of \citet{luenberger1969optimization} indicates that a necessary condition for a local maximum $\pi^*$ is that for all $\pi \in \Omega$, 
\[ \partial V(\pi; \pi - \pi^*)  \leq 0 \] 

Let $\rho(\pi, x) = f_x(\pi, x) \tau_x(\pi, x) + s(\pi, x)$. We can prove by contradiction that the optimal targeting policy must meet the conditions in the theorem. If there is some $\bar \pi$ that is optimal but does not meet the conditions in the theorem, then, one of the following must be true: 

\begin{enumerate} 
\item For $x$ in some set $Q$ that occur with non-zero probability, $\rho(\bar \pi, x) < 0$ but $\bar \pi(x) > 0$. But then choose $\pi$ such that $\pi(x) = \bar \pi(x)$ for $x \notin Q$ and $\pi(x) = 0$ for $x \in Q$. We have that  
\begin{equation*}
\begin{split}
 \partial V(\pi; \pi - \pi^*) = \int_{x \in Q} \rho(\bar \pi, x) ( 0 - \bar \pi(x)) > 0,
 \end{split} 
 \end{equation*} 
 which contradicts the optimality of $\bar \pi$.
 \item Or, for $x$ in some set $P$ that occurs with non-zero probability, $\rho( \bar \pi, x) > 0$ but $\bar \pi(x) < 1$. Choose $\pi$ such that $\pi(x) = \bar \pi(x)$ for $x \notin P$ and $\pi(x) = 1$ for $x \in P$. We have that 
 \begin{equation*}
\begin{split}
 \partial V(\pi; \pi - \pi^*) = \int_{x \in Q} \rho(\bar \pi, x) ( 1 - \bar \pi(x)) > 0,
 \end{split} 
 \end{equation*} 
 which contradicts the optimality of $\bar \pi$.
\end{enumerate} 
\end{proof} 

\subsection{Proof of Proposition \ref{prop:utility} } 
\label{pf:utility}
For the first part of the Proposition, we use Proposition \ref{prop:suf}. Since we have a binary covariate, we just need to show that $Pr(X_i^{\pi} = A | \tau_i = \tau)$ is differentiable in $\pi(A)$ and $\pi(B)$. Let $F^{1}_{c|\tau}(a|\tau)$ be the CDF of  $C_{iA} - C_{iB}$ conditional on $Z_i = 1$ and $\tau_i = \tau$.

\[ Pr(X_i^{\pi} = A | \tau_i = \tau) = Pr(u_k(B, \pi, \theta_i) \leq u_k(A, \pi, \theta_i)) \] 

For $k \in \{0, \ldots, 3\}$ we can write this as the following,  for some function $g(q, v)$ that is differentiable in both of its arguments. For example, for $u_0(x, \pi, \theta_i)$, then $g(q, v) = 0$. For $u_3(x, \pi, \theta_i)$, then $g(q, v) = q\cdot v - r\cdot q(1-q)$. 
\begin{equation*}
\begin{split} 
Pr(X_i^{\pi} = A | \tau_i = \tau)  &= \int  Pr\Big(C_{iA} - C_{iB} \leq g(\pi(A), v) - g(\pi(B), v) | \tau \Big) dF_v(v) \\
&= Pr(Z_i = 1) \int F^{1}_{c}\Big(g(\pi(A), v) - g(\pi(B), v)\Big |\tau\Big)dF_v(v)   \\ & \qquad +  Pr(Z_i = 0) Pr(X_i^{\pi} = A | \tau_i = \tau, Z_i = 0)\end{split}
\end{equation*} 

When we take the derivative with respect to $\pi(x)$, then only the term with $Z_i = 1$ has a non-zero derivative, since agents with $Z_i = 0$ never change their signal. The distribution function $F^{1}_{c|\tau}(a|\tau)$ has a derivative in $a$ since we assumed in the Proposition that the random variable $C_{iA} - C_{iB} | \tau_i = \tau, Z_i = 1$ has a density function, which we denote $f^{1}_{c|\tau}(a|\tau)$. Using the chain rule, we can now show that the conditional marginal distribution of $X_i^{\pi} = A$ is differentiable in $\pi(A)$ or $\pi(B)$. 

\begin{equation*}
\begin{split} 
& \frac{\partial Pr(X_i^{\pi} = A | \tau_i = \tau) }{\partial \pi(A)} = Pr(Z_i = 1) \int f^{1}_{c}\Big(g(\pi(A), v) - g(\pi(B), v)\Big | \tau \Big ) \frac{\partial g(\pi(A), v)}{ \partial \pi(A)} dF_v(v) \\
& \frac{\partial Pr(X_i^{\pi} = A | \tau_i = \tau) }{\partial \pi(B)} = -Pr(Z_i = 1) \int f^{1}_{c}\Big(g(\pi(A), v) - g(\pi(B), v)\Big | \tau \Big ) \frac{\partial g(\pi(B), v) }{ \partial \pi(B)} dF_v(v) \end{split}
\end{equation*}

For the counterexample in the proposition, imagine we have $C_{iA} = 0$ and $C_{iB} \sim \mbox{Uniform}(0, 1)$. Then, for any $\epsilon > 0$, when $\pi(B) = 0.5 - \epsilon$ and $\pi(A) = 0$, then $Pr(X_i^{\pi} = A) = 1$. But when $\pi(B) = 0.5 + \epsilon$, then we have $Pr(X_i^{\pi} = B) = 1$. The marginal distribution of $X_i^{\pi}$ does not vary smoothly in $\pi(x)$, since the utility function for every agent is discontinuous in $\pi(x)$ at the same value of 0.5. 

\subsection{Proof of Proposition \ref{prop:suf} } 
\label{pf:suf}
\begin{proof} 
When $X^{\pi}_i$ is discrete, we can write Equation \ref{eqn:first} as : 
\begin{equation*}
\begin{split} &  \lim \limits_{ \delta \rightarrow 0 } \int  \tau \cdot  \pi(x)  \frac{dF^{\pi + \delta h} _{x, \tau}(x, \tau)  - dF^{\pi}_{x, \tau} (x, \tau ) }{\delta} \\&= \lim \limits_{\delta \rightarrow 0}\int \tau \sum \limits_{x}  \pi(x) \frac{Pr(X_i^{\pi + \delta h} = x | \tau_i = \tau) - Pr(X_i^{\pi} = x| \tau_i = \tau)}{ \delta} dF_{\tau}(\tau)\\ 
& = \int \tau \sum \limits_{x}  \pi(x)  \lim \limits_{\delta \rightarrow 0} \frac{Pr(X_i^{\pi + \delta h} = x | \tau_i = \tau) - Pr(X_i^{\pi} = x | \tau_i = \tau)}{ \delta} dF_{\tau}(\tau) \\
& \stackrel{(1)}{=} \int \tau  \sum \limits_{x} \pi(x) \sum \limits_{x'} h(x') \frac{\partial Pr(X_i^{\pi} = x | \tau_i = \tau) }{\partial \pi(x')} dF_\tau(\tau)\\
& = \int \tau \sum \limits_{x'} h(x') \sum \limits_{x}  \pi(x) \frac{\partial Pr(X_i^{\pi} = x | \tau_i = \tau) }{\partial \pi(x')} dF_{\tau}(\tau) \\
& \stackrel{(2)}{=}\sum \limits_{x} h(x)  \int \tau \sum \limits_{x'}  \pi(x') \frac{\partial Pr(X_i^{\pi} = x' | \tau_i = \tau) }{\partial \pi(x)} dF_{\tau}(\tau) \\
& \stackrel{(3)}{=} \sum \limits_{x} h(x)  \sum \limits_{x'\neq x}   (\pi(x') - \pi(x))  \int \tau \frac{\partial Pr(X_i^{\pi} = x' | \tau_i = \tau) }{\partial \pi(x)} dF_{\tau}(\tau) \\
& = \sum \limits_{x} h(x) s(\pi, x) 
 \end{split}
\end{equation*} 

where $s(\pi, x) = \sum_{x' \neq x } [\pi(x') - \pi(x) ]\mathbb E \left [ \tau_i \cdot \frac{\partial Pr(X_i^{\pi} = x' | \tau_i)}{\partial \pi(x) } \right]$.
\begin{itemize} 
\item For (1), since we have that the marginal distribution of $X_i^{\pi}$ conditional on $\tau_i$ is differentiable in $\pi(x)$ for each $x \in \mathcal X$, we have that 
\[ \lim \limits_{\delta \rightarrow 0} \frac{Pr(X_i^{\pi + \delta h} = x | \tau_i = \tau) - Pr(X_i^{\pi} = x| \tau_i = \tau)}{ \delta} = \sum \limits_{x'} h(x') \frac{\partial Pr(X_i^{\pi} = x| \tau_i = \tau) }{\partial \pi(x')} \] 
\item For (2), the order of summation over $x$ and $x'$ is swapped. 
\item For (3), we rewrite by noticing $ Pr(X_i^{\pi} = x | \tau_i = \tau) = 1 - \sum \limits_{x' \neq x} Pr(X_i^{\pi} = x'| \tau_i = \tau)$. 
\end{itemize} 

\end{proof}

\subsection{Extension of Proposition \ref{prop:suf} to Continuous Covariates} 
\label{pf:ext}
We extend Proposition \ref{prop:suf} to the setting with continuous covariates. Define the conditional density function $f_{\pi}(x| \tau)$ using the CDF as $F_{\pi}(x | \tau)  = \int_{x} f_{\pi}(x | \tau) dx$. 
\begin{proposition} \label{prop:ext}
Let $X_i \in \mathbb R^d$. Assume that for each $\tau \in \mathcal T$,  the conditional density function $f^{\pi}_x(x | \tau)$ exists and is Gateaux-differentiable in $\pi$, with $\partial f^{\pi}_x(x | \tau; h) = \int h(x') g^{\pi}(x, x' | \tau) dx' $ for direction $h \in \Omega$. Then, Condition \ref{eqn:first} of Theorem \ref{thm:first} holds, with
\[  \lim \limits_{ \delta \rightarrow 0 } \int  \tau \cdot  \pi(x)  \frac{dF^{\pi + \delta h} _{x, \tau}(x, \tau)  - dF^{\pi}_{x, \tau} (x, \tau ) }{\delta} = \int h(x) s(\pi, x) d \mu (x),\] 
where 
\[ s(\pi, x) =\int_{x'} \pi(x') \mathbb E[\tau_i  g^{\pi}( x', x | \tau_i)]dx'.\] 
\end{proposition} 

We can write the integral over the joint distribution in terms of an integral over conditional distributions, and swap the limit and integral using the dominated convergence theorem. 
\begin{equation*} 
\begin{split} 
\lim \limits_{ \delta \rightarrow 0 } \int  \tau \cdot  \pi(x)  \frac{dF^{\pi + \delta h} _{x, \tau}(x, \tau)  - dF^{\pi}_{x, \tau} (x, \tau ) }{\delta} &= \lim \limits_{ \delta \rightarrow 0 } \int  \tau  \int  \pi(x) \frac{ f^{\pi + \delta h}(x| \tau) - f^{\pi }(x| \tau) dx }{\delta} d F_{\tau}(\tau)  \\
& = \int  \tau  \int \pi(x) \partial f^{\pi}_x(x | \tau; h)dx dF_{\tau}(\tau) \\ 
& \stackrel{(1)}{=} \mathbb E \left[\tau_i \int \pi(x)  \int h(x') g^{\pi}(x, x' | \tau_i) dx' dx \right ] \\
& = \mathbb E \left[\tau_i \int  h(x)  \int \pi(x') g^{\pi}(x', x | \tau_i) dx'   dx \right ] \\ 
& = \int h(x) \mathbb E \left[\tau_i     \int \pi(x') g^{\pi}(x', x | \tau_i) dx'  \right ]  dx \\ 
& = \int h(x) s(\pi, x) dx
\end{split} 
\end{equation*} 

In (1), we plug in the assumed form for $\partial f^{\pi}_x(x | \tau; h) $. The remaining steps swap the order of integration until we have the desired form for the directional derivative of $V(\pi)$. 

\subsection{Differentiability Condition in Proposition \ref{prop:suf} is Not Necessary} 
\label{pf:comment} 

We introduce a simple model with binary covariates where the marginal distribution of $X_{i}^{\pi} | \tau_i = \tau$ is not differentiable in $\tau$. In this model $\theta_i = \{Y_i(1), Y_i(0)\}$. $Y_i(0) = 0$ and $\tau_i \sim \mbox{Uniform}(-2, 2)$. In this model 

\[ x(\pi, \theta_i) = B + A \mathbbm{1}( \pi(A) - \pi(B) \geq \tau_i). \] 

In this model, $Pr(X_i^{\pi} = A | \tau_i = \tau) = \mathbbm{1}( \pi(A) - \pi(B) \geq \tau_i )$ is not differentiable in $\pi(A)$ and $\pi(B)$, as Proposition \ref{prop:suf} requires. However, we can still derive an $s(\pi, x)$ such that Condition \ref{eqn:first} in Theorem \ref{thm:first} still holds. We can write 

\begin{equation*} 
\begin{split}
V(\pi) & = \int_{-2}^2 \tau \cdot \pi(A) \mathbbm{1}( \pi(A) - \pi(B) \geq \tau)  + \tau \cdot \pi(B) \mathbbm{1}( \pi(A) - \pi(B) < \tau) dF_{\tau}(\tau) \\ 
&=   \int_{-2}^{\pi(A)- \pi(B)} \pi(A) \tau dF_{\tau}(\tau) + \int_{\pi(A) - \pi(B)}^2 \pi(B)  \tau dF_{\tau}(\tau) \\ 
\end{split} 
\end{equation*} 

We can now derive $s(\pi, x)$: 

 \begin{equation*} 
 \begin{split} \lim \limits_{ \delta \rightarrow 0 } \int  \tau \cdot  \pi(x)  \frac{dF^{\pi + \delta h} _{x, \tau}(x, \tau)  - dF^{\pi}_{x, \tau} (x, \tau ) }{\delta} & = \lim \limits_{\delta \rightarrow 0}  \int_{-2}^2 \tau \cdot \pi(A) \mathbbm{1}( \pi(A) + \delta h(A) - \pi(B) - \delta h(B) \geq \tau) dF_{\tau}(\tau) \\ & \qquad + \lim \limits_{\delta \rightarrow 0} \int_{-2}^{2} \tau \cdot \pi(B) \mathbbm{1}( \pi(A) + \delta h(A) - \pi(B) - \delta h(B) < \tau) dF_{\tau}(\tau)  \\ 
 & = \lim \limits_{\delta \rightarrow 0} \int_{-2}^{\pi(A) + \delta h(A)- \pi(B) - \delta h(B)} \pi(A) \tau dF_{\tau}(\tau) \\ & \qquad + \lim \limits_{\delta \rightarrow 0}\int_{\pi(A) + \delta h(A) - \pi(B) - \delta h(B)}^2 \pi(B)  \tau dF_{\tau}(\tau)\\ 
 & = ( \pi(A)  - \pi(B) (\pi(A) - \pi(B))f_{\tau}(\pi(A) - \pi(B)) \cdot h(A) \\ & \qquad - (\pi(A) - \pi(B))^2 f_{\tau}(\pi(A) - \pi(B)) \cdot h(B), 
 \end{split} 
 \end{equation*} 

where $f_{\tau}(\tau)$ is the density of $\tau_i$. We took the derivative using the Leibniz integral rule. This derivation shows that $s(\pi, A) =  ( \pi(A)  - \pi(B))^2 f_{\tau}(\pi(A) - \pi(B)) $ and $s(\pi, B) =  -( \pi(A)  - \pi(B))^2 f_{\tau}(\pi(A) - \pi(B)) $. This proves that the differentiability condition in Proposition \ref{prop:suf} is not necessary for Condition \ref{eqn:first} of Theorem \ref{thm:first} to hold. 

\subsection{Proof of Corollary \ref{cor:tilde}} 

To avoid a double superscript, for this proof we use the notation $s(\pi, x) = s(\pi, x)$ and $f_x(\pi, x) = f_x(\pi, x)$. We proceed by contradiction. Let's assume that there is some $\tilde x \in \mathcal X$ with $\mbox{sgn}( s(\pi^c, \tilde x)) \neq \mbox{sgn} ( \tau(\tilde x, \pi^c))$ and $|s (\pi^c, \tilde x)| > | f(\pi^c, x) \tau(\pi^c, \tilde x) |$ but $\pi^c$ is still optimal. 

$\pi^c(\tilde x)$ is either 0 or 1. If it is 1, then $\tau(\pi^c, x)> 0$. But then by definition of $\tilde x$, we have that $s(\pi^c, \tilde x) + \tau(\pi^c, \tilde x) < 0$. By Theorem \ref{thm:first}, no optimal rule can have both $\pi^c(\tilde x) = 1$ and $s(\pi^c, \tilde x) + f(\pi^c, x)  \tau(\pi^c, \tilde x) < 0$. So it can't be that $\pi^c$ is optimal. 

A similar argument holds if $\pi^c(\tilde x) = 0$. By definition, if it is 0, then $\tau(\pi^c, x)< 0$. By the definition of $\tilde x$, we have that  $s(\pi^c, \tilde x) + f(\pi^c, x)  \tau(\pi^c, \tilde x) > 0$. But by Theorem \ref{thm:first}, no optimal rule can have both $\pi^c(\tilde x) = 0$ and $s(\pi^c, \tilde x) + \tau(\pi^c, \tilde x) > 0$. We have now shown that if $\tilde x$ with the property given in the Corollary exists, then a cutoff rule is not optimal. 

If no such $\tilde x$ exists, then for every $x \in \mathcal X$, we have that $\pi^c(x) = 0$ and $s(\pi^c, \tilde x) + f(\pi^c, x)  \tau(\pi^c, \tilde x) \leq 0$ or $\pi^c(x) = 1$ and $s(\pi^c, \tilde x) + f(\pi^c, x)  \tau(\pi^c, \tilde x) \geq 0$. Then for every $x \in \mathcal X$, $\pi^c(x)$ meets the necessary conditions for optimality given in Theorem \ref{thm:first}. If $V(\pi)$ is concave, then the necessary conditions from Theorem \ref{thm:first} are also sufficient and this ensures that $\pi^c(x)$ is not just locally optimal, but also globally optimal.

\subsection{Verifying Assumption \ref{ass:strat} Models Meet Condition \ref{eqn:first} of Theorem \ref{thm:first}} 
\label{pf:binary} 

We can use Proposition \ref{prop:suf}, which expresses $s(\pi, x)$ in terms of the derivative of the marginal distribution of $X_{i}^{\pi}$  conditional on $\tau_i$ with respect to $\pi(x)$. 

$Pr(X_i^{\pi} = A | \tau_i  = M ) = Pr(C_{iA} \leq V_M (\pi(A) - \pi(B)))$. Let $f^M_{cA}$ be the density of $C_{iA}$ for agents with $\tau_i = M$. 

\begin{equation*} 
\begin{split}
& \frac{\partial Pr(X_i^{\pi} = B | \tau_i  = M )}{ \partial \pi(A)} = - V_M f^M_{cA}\Big(V_M[\pi(A) - \pi(B)]\Big) \\
& \frac{\partial Pr(X_i^{\pi} = B | \tau_i  = M )}{ \partial \pi(B)} =  V_M f^M_{cA}\Big (V_M[\pi(A) - \pi(B)] \Big)  
\end{split} 
\end{equation*} 
From Proposition \ref{prop:suf}, this implies that Condition \ref{eqn:first} of Theorem \ref{thm:first} holds for any model that meets Assumption \ref{ass:strat}. Given we have the derivatives of the conditional marginal distribution,  from Proposition \ref{prop:suf} we can derive
\begin{equation*} 
\begin{split} 
 s(\pi, A) & = \sum_{x' \neq A } [\pi(x') - \pi(A) ]\mathbb E \left [ \tau_i \cdot \frac{\partial Pr(X_i^{\pi} = x' | \tau_i)}{\partial \pi(A) } \right]   \\ 
 & = [\pi(A) - \pi(B) ]\cdot  \rho_M \cdot  M \cdot V_M \cdot f^M_{cA}\Big(V_M[\pi(A) - \pi(B) ]\Big)
\end{split}
\end{equation*} 

\subsection{Proof of Proposition \ref{prop:dro}}
\label{pf:dro} 

\begin{proof} 
We can prove this by contradiction. Assume we have $x \in \mathcal X$ with $\tau_x(\pi, x) > 0$ for all $\pi \in \Omega$, but $\pi^R(x) < 1$. Let $ \tau_{MIN} = \min \limits_{F \in \mathcal Q} Pr(X_i^{\pi} = x) \mathbb E[Y_i(1) - Y_i(0) | X^{\pi}_i = x]$, where $Q = \{ F^{\pi}_{x, \tau} | \pi \in \Omega \}$.  $\tau_{MIN} > 0$. Let $\pi'(x') = \pi^R(x')$ for $x' \neq x$, and $\pi'(x) = 1$. 

\begin{equation*} 
\begin{split} 
 &\min_{F \in \mathcal Q} \mathbb E_F[\pi^R(X_i) Y_i(1) - Y_i(0)] - \mathbb E_F[\pi'(X_i) Y_i(1) - Y_i(0)] \\ 
= &  \tau_{MIN} (\pi^R(X_i) - \pi'(X_i))
< 0 
\end{split} 
\end{equation*} 
This contradicts the min-max optimality of $\pi^R$. 
Assume we have $x \in \mathcal X$ with $\tau_x(\pi, x) < 0$ for all $\pi \in \Omega$, but $\pi^R(x) > 0$. Let $ \tau_{MIN} = \min \limits_{F \in \mathcal Q}  f_{\pi} (x) \mathbb E[Y_i(1) - Y_i(0) | X^{\pi}_i = x]$, where $Q = \{ F^{\pi}_{x, \tau} | \pi \in \Omega \}$. We have  $ \tau_{MIN} < 0$. Let $\pi'(x') = \pi^R(x)$ for $x' \neq x$, but $\pi'(x) = 0$. 

\begin{equation*} 
\begin{split} 
 &\min_{F \in \mathcal Q} \mathbb E_F[\pi^R(X_i) Y_i(1) - Y_i(0)] - \mathbb E_F[\pi'(X_i) Y_i(1) - Y_i(0)] \\ 
= &   \tau_{MIN} (\pi^R(X_i) - \pi'(X_i))
< 0 
\end{split} 
\end{equation*} 

This contradicts the min-max optimality of $\pi^R$. So, $\pi^R$ must take the form of the cutoff rule in Proposition \ref{prop:dro}. 
\end{proof}

\section{Solving Example Models} 

\subsection{Solving Coupon Allocation Model} 
\label{app:coupon} 

We can write the objective function as 
\begin{equation*} 
\begin{split}
V(\pi) & = \pi(A) [Pr(X_i = A | \gamma_i = 1)Pr(\gamma_i = 1)Y(1 , \gamma_i = 1) + Pr(X_i = A | \gamma_i = 0)Pr(\gamma_i = 0)Y(1 , \gamma_i = 0) \\ 
& + (1 - \pi(A))  [Pr(X_i = A | \gamma_i = 1)Pr(\gamma_i = 1)Y(0 , \gamma_i = 1) + Pr(X_i = A | \gamma_i = 0)Pr(\gamma_i = 0)Y(0 , \gamma_i = 0) ]\\ 
& +  \pi(B) [Pr(X_i = B | \gamma_i = 1)Pr(\gamma_i = 1)Y(1 , \gamma_i = 1) + Pr(X_i = B | \gamma_i = 0)Pr(\gamma_i = 0)Y(1 , \gamma_i = 0) ] \\
& + ( 1- \pi(B)) [Pr(X_i = B | \gamma_i = 1)Pr(\gamma_i = 1)Y(0, \gamma_i = 1) + Pr(X_i = B | \gamma_i = 0)Pr(\gamma_i = 0)Y(0, \gamma_i = 0) ]
\end{split} 
\end{equation*} 
$Y(1, \gamma_i = 1) = 3.75$, $Y(1, \gamma_i = 0) = 5$, $Y(0, \gamma_i = 1) = 0$, and $Y(1, \gamma_i = 0)$ = 10. $Pr(X_i = A | \gamma_i = 1) = 1$, so we just have to derive $Pr(X_i = A | \gamma_i = 0)$. 
\begin{equation*}
\begin{split} 
Pr(X_i = A | \gamma_i = 0) &= Pr\big(C_i \leq 5 (\pi(A) - \pi(B))\big) \\ 
& = \frac{1}{2}(\pi(A) - \pi(B) ) 
\end{split} 
\end{equation*} 

We can then plug into the objective, 

\begin{equation*} 
\begin{split}
V(\pi) & = \pi(A) \Big [1.875 + \frac{5}{4} (\pi(A) - \pi(B))\Big] + ( 1 - \pi(A))\frac{10}{4}(\pi(A) - \pi(B)) \\
& + \pi(B)\frac{5}{4}(2 - \pi(A) + \pi(B)) + ( 1- \pi(B)) \frac{10}{4} (2 - \pi(A) + \pi(B))
\end{split} 
\end{equation*} 

which can be simplified to 
\[ V(\pi) =  -0.625(2\pi(B)^2 -4 \pi(B) \pi(A) + 4 \pi(B) + 2\pi(A)^2 - 3 \pi(A)  -8) \] 

Note that this objective is concave. 
We can find the global optimum using the KKT conditions, where we constraint $0 \leq \pi(A) \leq 1$, and $0 \leq \pi(B) \leq 1$
For $\pi(B)$, we have: 

\[ -\frac{5}{2} (\pi(B) - \pi(A) +1)  + \lambda^0_L = 0 \]

For $\pi(A)$, we have: 
\[ 2.5 (\pi(B) - \pi(A) + 0.75) - \lambda^1_H  = 0 \]

We can't have both $\pi(B)$ and $\pi(A)$ be interior solutions. There is also no solution satisfying the KKT conditions when $\pi(A) = 1$. The solution to the KKT conditions has $\pi^*(B) = 0$ and $\pi^*(A) = 0.75$.







\subsection{Solving Product Upgrade Model} 
\label{app:upgrade} 

As in the previous example, we can write the objective function as
\begin{equation*} 
\begin{split}
V(\pi) & = \pi(A) [Pr(X_i = A | \gamma_i = 1)Pr(\gamma_i = 1)Y(1 , \gamma_i = 1) + Pr(X_i = A | \gamma_i = 0)Pr(\gamma_i = 0)Y(1 , \gamma_i = 0) \\ 
& + (1 - \pi(A))  [Pr(X_i = A | \gamma_i = 1)Pr(\gamma_i = 1)Y(0 , \gamma_i = 1) + Pr(X_i = A | \gamma_i = 0)Pr(\gamma_i = 0)Y(0 , \gamma_i = 0) ]\\ 
& +  \pi(B) [Pr(X_i = B | \gamma_i = 1)Pr(\gamma_i = 1)Y(1 , \gamma_i = 1) + Pr(X_i = B | \gamma_i = 0)Pr(\gamma_i = 0)Y(1 , \gamma_i = 0) ] \\
& + ( 1- \pi(B)) [Pr(X_i = B | \gamma_i = 1)Pr(\gamma_i = 1)Y(0, \gamma_i = 1) + Pr(X_i = B | \gamma_i = 0)Pr(\gamma_i = 0)Y(0, \gamma_i = 0) ]
\end{split} 
\end{equation*} 

$Y(1, \gamma_i = 1) = 10$, $Y(1, \gamma_i = 0) = 1$, $Y(0, \gamma_i = 1) = 5$, and $Y(0, \gamma_i = 0)$ = 5. $Pr(X_i = B | \gamma_i = 0) = 1$, so we just have to derive $Pr(X_i = A | \gamma_i = 1)$. 
\begin{equation*}
\begin{split} 
Pr(X_i = A | \gamma_i = 1) &= Pr\big(-V_i \leq 5 (\pi (H) - \pi(B))\big) \\ 
& = \frac{1}{4}(\pi(A) - \pi(B) ) + \frac{1}{2} 
\end{split} 
\end{equation*} 

We can then plug into the objective, 

\begin{equation*} 
\begin{split}
V(\pi) & = \pi(A) \Big [\frac{5}{4} (\pi(A) - \pi(B)) + \frac{5}{2} \Big] + ( 1 - \pi(A))\Big[\frac{5}{8}(\pi(A) - \pi(B)) + \frac{5}{4} \Big ]\\
& + \pi(B)( 3 - \frac{5}{4} (\pi(A) - \pi(B)) + ( 1- \pi(B)) \Big[\frac{15}{4} - \frac{5}{8}( \pi (H) - \pi(B)) \Big]
\end{split} 
\end{equation*} 
which can be simplified to 

\[ 
V(\pi) = \frac{1}{8}[ 5 \pi(B)^2 - 10 \pi(B) \pi(A) - 6 \pi(B) + 5 \pi(A)^2 + 10 \pi(A) + 40) 
\]
Note that this objective is concave and we can find the global optimum using KKT conditions: 

For $\pi(B)$, we have: 

\[ \frac{1}{4}(5 \pi(B) - 5 \pi(A) -3 ) + \lambda^0_B = 0 \]

For $\pi(A)$, we have: 
\[ \frac{-5}{4} (\pi(B) - \pi(A) - 1)  - \lambda^1_A= 0 \]

The KKT conditions are satisfied with $\pi^*(B) = 0$ and $\pi^*(A) = 1$ at which the objective value is 6.875. 





\section{MTurk Survey Details} 
\label{app:mturk} 
\subsection{Survey Questions} 

Bonus and no-incentive survey versions were identical except for Question 4. 

\paragraph{Question 1} (Multiple Choice) \textit{How much do you like math, on a scale of 1 to 5? }
\begin{itemize} 
\item 1 (I strongly dislike math) 
\item 2 (I dislike math) 
\item 3 (I feel neutral towards math) 
\item 4 (I like math) 
\item 5 (I love math) 
\end{itemize} 

\paragraph{Question 2} (Multiple Choice) \textit{How proficient do you consider yourself at math on a scale of 1 (beginner) to 5 (expert)? }
\begin{itemize} 
\item 1 
\item 2 
\item 3
\item 4 
\item 5
\end{itemize} 

\paragraph{Question 3} (Multiple Choice) \textit{What is your education level? }
\begin{itemize} 
\item Less than high school 
\item High School graduate
\item Less than 2 years of college
\item 2 years of college
\item 4 years of college
\item 2 year masters/professional degree
\item 4+ year doctoral/professional degree
\end{itemize} 

\paragraph{Question 4 for No-Incentive Survey}  (Text Entry) \textit{ (Optional, but encouraged to attempt). $2x^2 - 5x = 3$. What is $x$? If there are multiple correct answers, choose one.} 

\paragraph{Question 4 for Bonus Survey} (Text Entry). \textit{(Optional, but encouraged to attempt). As an incentive to attempt this question, we will pay a bonus of \$1 for correctly answering this question. $2x^2 - 5x = 3$. What is $x$? If there are multiple correct answers, choose one. }

\subsection{HIT Description} 

We are conducting a research project titled “Experimental Design of Robust Predictive Policy”. The survey consists of three simple questions on demographics and opinions and a skill-testing question. The information that you report will be used to test the performance of an algorithm for predicting outcomes based on self-reported data. The fixed payment of \$0.12 may be augmented by a small bonus in some versions of the survey, depending on your response to one of the questions. 

Your participation will take approximately one to three minutes.  

Your participation is voluntary and you have the right to withdraw your consent or discontinue participation at any time without penalty or loss of benefits to which you are otherwise entitled.

By completing the survey, you consent for our team to use your submitted information to evaluate our prediction algorithm. If you agree to participate in this research, please complete the survey by selecting the link below. At the end of the survey, you will receive a code to paste into the box below to receive credit for taking our survey.

\textbf{Make sure to leave this window open as you complete the survey.} When you are finished, you will return to this page to paste the code into the box.

\end{document}